\documentclass[pre,aps,amsmath,amssymb,amsfonts,floatfix,superscriptaddress,showpacs,twocolumn,footinbib]{revtex4-1}
\usepackage{graphicx}
\usepackage{color}
\usepackage[utf8]{inputenc}
\usepackage{dcolumn}
\usepackage{hyperref}
\usepackage{textcomp}
\usepackage{enumitem}

\usepackage{comment}
\usepackage[dvipsnames]{xcolor}
\usepackage{tikz}
\usetikzlibrary{calc}
\usetikzlibrary{decorations.pathmorphing}
\usetikzlibrary{decorations.pathreplacing}
\usetikzlibrary{decorations.markings}
\usetikzlibrary{shapes.misc}
\usetikzlibrary{positioning}
\usetikzlibrary{snakes}
\usetikzlibrary{3d}
\def\checkmark{\tikz\fill[scale=0.4,color=red](0,.35) -- (.25,0) -- (1,.6) -- (.25,.15) -- cycle;}

\hypersetup{colorlinks=true,breaklinks,linkcolor=blue,urlcolor=blue,citecolor=blue}

\newcommand{\vect}[1]{\boldsymbol{#1}}

\newcommand{\kv}{\ensuremath{\mathbf{k}}}
\newcommand{\qv}{\ensuremath{\mathbf{q}}}

\newcommand{\av}[1]{\ensuremath{\left\langle #1 \right\rangle}}
\newcommand{\up}{\ensuremath{\uparrow}}
\newcommand{\dn}{\ensuremath{\downarrow}}

\newcommand{\vc}[1]{\ensuremath{\mathbf{#1}}}

\begin{document}

\title{Conservation in two-particle self-consistent extensions of dynamical-mean-field-theory}

\author{Friedrich Krien}
\affiliation{Institute of Theoretical Physics, University of Hamburg, 20355 Hamburg, Germany}

\author{Erik G. C. P. van Loon}
\affiliation{Radboud University, Institute for Molecules and Materials, NL-6525 AJ Nijmegen, The Netherlands}
  
\author{Hartmut Hafermann}
\affiliation{Mathematical and Algorithmic Sciences Lab, Paris Research Center, Huawei Technologies France SASU, 92100 Boulogne Billancourt, France}

\author{Junya Otsuki}
\affiliation{Department of Physics, Tohoku University, Sendai 980-8578, Japan}

\author{Mikhail I. Katsnelson}
\affiliation{Radboud University, Institute for Molecules and Materials, NL-6525 AJ Nijmegen, The Netherlands}

\author{Alexander I. Lichtenstein}
\affiliation{Institute of Theoretical Physics, University of Hamburg, 20355 Hamburg, Germany}

\setcounter{page}{1}

\begin{abstract}
Extensions of dynamical-mean-field-theory (DMFT) make use of quantum impurity models as non-perturbative and exactly solvable reference systems which are essential to treat the strong electronic correlations.
Through the introduction of retarded interactions on the impurity, these approximations can be made two-particle self-consistent.
This is of interest for the Hubbard model, because it allows to suppress the antiferromagnetic phase transition in two-dimensions in accordance with the Mermin-Wagner theorem,
and to include the effects of bosonic fluctuations. For a physically sound description of the latter, the approximation should be conserving.
In this paper we show that the mutual requirements of two-particle self-consistency and conservation lead to fundamental problems.
For an approximation that is two-particle self-consistent in the charge- and longitudinal spin channel, the double occupancy of the lattice and the impurity are no longer consistent when computed from single-particle properties.
For the case of self-consistency in the charge- and longitudinal as well as transversal spin channels, these requirements are even mutually exclusive so that no conserving  approximation can exist.
We illustrate these findings for a two-particle self-consistent and conserving DMFT approximation.
\end{abstract}

\date{\today}

\maketitle

Models for correlated electron systems, such as the Hubbard model, count among the hardest problems of contemporary condensed matter physics. At the same time, they are believed to capture the physics of fascinating phenomena such as high-temperature superconductivity~\cite{Dagotto94} and the Mott transition~\cite{Imada98}. To understand the underlying physics, it is necessary to develop methods which can capture these phenomena. Because of unavoidable approximations however, it is not always possible to separate the physics from artifacts of the method. It is therefore desirable to design methods which satisfy basic requirements such as translational invariance, thermodynamic consistency~\cite{vanLoon15,Aichhorn06}, local conservation laws of charge and spin~\cite{Baym61,Baym62,Hafermann14-2} and, in view of an application to high-temperature superconductivity of layered Cuprates~\cite{Dagotto94}, the Mermin-Wagner theorem~\cite{Mermin1966}.

Dynamical mean-field theory (DMFT)~\cite{Georges96} and its cluster extensions~\cite{Maier05} have been an important step towards the understanding of correlated electron behavior, in particular the Mott transition.
From a theoretical perspective, DMFT can be thought of as an approximation to the exact Luttinger-Ward functional, where all propagators are replaced by the corresponding local ones. An auxiliary problem subject to a self-consistency condition --often an Anderson impurity model (AIM)-- is used as a tool to sum the diagrams of this local functional exactly.  As a consequence, DMFT is conserving in the Baym-Kadanoff sense~\cite{Georges96,Potthoff06-2,Hafermann14-2,Baym62}. 

The more recently introduced diagrammatic extensions of DMFT such as D$\Gamma$A~\cite{Toschi07,Rohringer16}, the dual fermion (DF)~\cite{Rubtsov08}, one-particle irreducible (PI)~\cite{Rohringer13}, TRILEX~\cite{Ayral15,Ayral15-2}, DMF${}^2$RG~\cite{Taranto14}, and dual boson (DB) approaches~\cite{Rubtsov12} are an active field of research.
The AIM plays a central role in these approaches. From a suitable dynamical vertex function of the AIM non-local approximations to the self-energy are constructed by summing certain classes of diagrams. The lattice self-energy is hence approximate, but incorporates long-range correlations. The use of dynamical vertices allows one to deal with strong correlations as opposed to approaches based on the bare Hubbard interaction, such as the fluctuation exchange approximation (FLEX)~\cite{Bickers89} or the two-particle self-consistent approach (TPSC)~\cite{Vilk97}.
Despite significant progress in this field~\cite{Rohringer2017}, a number of open questions remain. For example, it is not always clear how to choose the diagrams~\cite{Gukelberger2016,Iskakov16} or self-consistency conditions. More generally, the question is how to optimally exploit the AIM, whose solution we know (numerically) exactly, to construct approximations that meet the above mentioned basic requirements.

Extended dynamical mean-field theory (EDMFT)~\cite{Sengupta95,Otsuki13-2,sachdev93,Si96,Kajueter96,Smith00,Chitra01} includes the effect of two-particle bosonic fluctuations through a local retarded interaction in the impurity model that is fixed by a corresponding self-consistency condition. As a result, EDMFT, as well as its extensions, like (E)DMFT+$GW$~\cite{sun02,Ayral12}, are two-particle self-consistent.
A consequence of the EDMFT self-consistency condition is that the lattice double occupancy equals that of the impurity model and hence is bounded.
As Vilk and Tremblay have shown~\cite{Vilk97}, any approximation which produces a bounded lattice double occupancy will suppress magnetic phase transitions in two dimensions, as required by the Mermin-Wagner theorem.
This property is indeed respected by the spin-DMFT~\cite{Otsuki13-2} (which is akin to EDMFT), while it is violated in DMFT.

A disadvantage of EDMFT is, however, that it breaks Ward identities~\cite{Hafermann14-2,Otsuki13-2} and therefore violates local conservation laws.
This can lead to a qualitatively wrong description of the physics of the collective excitations.
For example, in presence of a long-range interaction, the energy of the plasmon diverges in the long wavelength limit.
In case of a local interaction, one obtains a plasmon-like feature instead of a zero-sound mode~\cite{Hafermann14-2}.

In the dual boson (DB) approach~\cite{Rubtsov12,vanloon14-2}, which is a diagrammatic extension of EDMFT, conservation in the charge channel can be restored by including certain ladder diagrams into the bosonic propagator. This provides a physically sound description of plasmons even in the correlated state~\cite{vanLoon14}.
Remarkably, global charge conservation is maintained in a two-particle self-consistent version of the approach~\cite{Stepanov16}.
The two-particle self-consistent theory also resolves an ambiguity in the computation of the double occupancy present in DMFT and yields results closer to benchmarks than either of the two DMFT values~\cite{vanLoon16}.

In particular for an application to superconductivity, it is desirable to include spin fluctuations as well, while maintaining the conserving character of the theory and a sound description of the collective modes.
Self-consistent approaches based on an impurity model including retarded spin-spin interactions have been considered previously~\cite{Sengupta95,Otsuki13-2,sachdev93}.
It seems appealing to include diagrammatic corrections in order to make such a theory conserving.

These considerations lead us to the following questions: Is it possible to construct a two-particle self-consistent version of DMFT which would be conserving and satisfy the Mermin-Wagner theorem?
Similarly, under which conditions can we extend EDMFT and spin-DMFT to satisfy the conservation laws?
Quantum impurity models are at the heart of these approaches and serve as exactly solvable reference systems~\cite{CTQMCRMP}.
They allow to treat at least part of the strong electronic correlations in a non-perturbative manner.
Hence more generally, the question is whether it is possible to construct approximations that exploit the non-perturbative starting point provided by the impurity model while maintaining the desirable properties mentioned above.

The aim of this paper is to show that one faces fundamental difficulties in an attempt to construct such approximations.
In particular, we demonstrate that for a conserving approximation, imposing two-particle self-consistency in the charge and one of the spin channels leads to an inconsistency in the calculation of the potential energy due to the retarded interactions.
More importantly, we prove that if one attempts to impose self-consistency in the charge and all three spin channels, no conserving approximation can exist.
In essence, we find that the retarded spin interactions, introduced to make the theory two-particle self-consistent, undermine the desired feature of local conservation. 
We show that this limitation is rooted in the fact that the Ward identities of the lattice and of the impurity model are incompatible.
As a concrete example, we construct a two-particle self-consistent DMFT which is conserving in the charge- and one of the spin-channels.

The paper is organized as follows:
We recollect the DMFT approximation to the Hubbard model in section~\ref{sec:DMFT} and examine the thermodynamic consistency of the total energy in this approximation.
We introduce two-particle self-consistency in section~\ref{sec:2psc} and perform a similar analysis.
A conflict between two-particle self-consistency and local conservation is related to the Ward identities of the impurity model in section~\ref{section:conservation_scdb}.
We present an application of a two-particle self-consistent DMFT in Sec.~\ref{section:scdb}.
We interpret our main results in Sec.~\ref{sec:discussion} and finally conclude in Sec.~\ref{sec:conclusions}.
Derivations for several analytical results are provided in the Appendices~\ref{app:Wardlat}-\ref{app:Wardlocal}.

\section{Dynamical mean-field theory}
\label{sec:DMFT}

To set the stage, we first discuss the familiar case of DMFT.
For concreteness, we focus on the two-dimensional (2D) paramagnetic Hubbard model on the square lattice with nearest-neighbor hopping given by the Hamiltonian
\begin{align}
\label{hubbardmodel}
H = &-t\sum_{\langle ij\rangle\sigma} c^\dagger_{i\sigma}c^{}_{j\sigma}+ U\sum_{i} n_{i\up} n_{i\dn}.
\end{align}
Here $i,j$ label lattice sites. The local Hubbard interaction has strength $U$. We use the hopping $t=1$ as the unit of energy and denote Green's function as $G_{ij}$ in real space and $G_{\kv}$ in momentum space respectively (when it is not ambiguous, we omit the frequency dependence for brevity).

DMFT is a local approximation to the exact Luttinger-Ward functional $\Phi[G_{ij}]\approx\sum_{i}\phi[G_{\text{loc}}]$, which is therefore conserving in the Baym-Kadanoff sense~\cite{Georges96,Baym62}.
As a result of the local approximation, the self-energy is local: $\Sigma_{ij}=\delta\Phi[G_{i'j'}]/\delta G_{ji}=\delta\phi[G_{\text{loc}}]/\delta G_{\text{loc}}\, \delta_{ji}$ and we note that the same holds for the irreducible vertex: $-\Gamma_{ijkl}=\delta^2\Phi[G_{i'j'}]/\delta G_{ji}\delta G_{lk}=\delta^2\phi[G_{\text{loc}}]/\delta G_{\text{loc}}^{2}\, \delta_{li}\delta_{lj}\delta_{lk}$~\footnote{A minus sign arises in the functional relation of $\Gamma$ due to the definitions chosen in this publication.}.
If we know the local Green's function, the problem is solved: In this case we can evaluate the local functional and its derivatives at the local Green's function and hence compute the self-energy.
Since we do not know the local Green's function a priori and the self-energy is a functional of the latter, we have to solve this problem self-consistently:
we vary $G_{\text{loc}}$ until the local Green's function computed from the self-energy equals $G_{\text{loc}}$.

We can employ an auxiliary local model as a \emph{tool} to accomplish this and to sum the diagrams of this local functional exactly.
This means letting $\phi[G_{\text{loc}}]\equiv \phi_{\text{imp}}[g_{\text{imp}}]$ and $\Sigma[G_{\text{loc}}]\equiv \Sigma_{\text{imp}}[g_{\text{imp}}]$.
The desired solution is evidently obtained when the DMFT self-consistency condition is satisfied,
\begin{align}
g_{\text{imp},\nu}=G_{\text{loc},\nu}.
\label{eq:DMFTsc}
\end{align}
(Where unambiguous, we drop labels 'imp' and 'lat' in what follows.)

In practice, an Anderson impurity model (AIM) is often employed for this purpose, whose action reads
\begin{align}
  \hspace*{-0.18cm}S_{\text{AIM}}=&\hspace*{-0.01cm}-\hspace*{-0.01cm}\sum_{\nu\sigma}c^*_{\nu\sigma}(\imath\nu+\mu-\Delta_\nu)c^{}_{\nu\sigma}+\hspace*{-0.01cm}U\sum_{\omega}n_{-\omega\uparrow} n_{\omega\downarrow}.
  \label{eq:AIM}
\end{align}
Here $\Delta_{\nu}$ denotes the electronic hybridization, $\mu$ is the chemical potential and $\nu$ ($\omega$) denote the discrete fermionic (bosonic) Matsubara frequencies $\nu_n=(2n+1)\pi/\beta$ and $\omega_m=2m\pi/\beta$, respectively. $\beta=1/T$ is the inverse temperature. The AIM has the same local interaction $U$ as the lattice model.

Let us now take a \emph{practical} viewpoint.
Assume we have a non-trivial model that we can solve exactly, such as the AIM described by the action~\eqref{eq:AIM}. From this model we can obtain the local impurity self-energy and irreducible vertex function. We can now ask the question of how to construct a conserving approximation given these quantities. 

We recall that \emph{local} conservation of charge and spin means that the following continuity equations for the charge ($\rho^{0}$) and spin densities ($\rho^{x,y,z}$) hold: 
\begin{align}
\partial_\tau\rho^\alpha=-[\rho^\alpha,H].
\label{eq:conteq}
\end{align}
We have introduced the index $\alpha=0,x,y,z$ to label the charge and spin channels.
The corresponding charge and spin density operators are defined as $\rho^\alpha = \sum_{\sigma\sigma'}c^\dagger_{\sigma}s^\alpha_{\sigma\sigma'}c_{\sigma'}$ with the Pauli matrices $s^\alpha$,
such that $\rho^0=n=n_\uparrow+n_\downarrow$ and $\rho^{x,y,z}=2S^{x,y,z}$.

On the lattice we can formulate the following Ward identities (cf. Appendix~\ref{app:Wardlat}), which are the Green's function analogues of the continuity equations~\eqref{eq:conteq}:
\begin{align}
\Sigma_{k+q}-\Sigma_{k}=-\sum_{k'}\Gamma^{\alpha}_{kk'q}[G_{k'+q}-G_{k'}].
\label{eq:Ward}
\end{align}
Here we have introduced four-vector notation $k\equiv(\kv,\nu)$ and $q\equiv(\qv,\omega)$. Summations over frequencies and momenta imply factors $\beta^{-1}$ and $N^{-1}$, respectively, with $N$ being the number of sites.
$\Sigma$ and $G$ are the exact lattice self-energy and Green's function, respectively, and $\Gamma^{\alpha}$ denotes the \textit{irreducible} (horizontal) \textit{particle-hole} vertex.
The irreducible vertices in the charge and spin channels are explicitly defined as $\Gamma^{0}=\Gamma^{\uparrow\uparrow\uparrow\uparrow}+\Gamma^{\uparrow\uparrow\downarrow\downarrow}$,
$\Gamma^{z}=\Gamma^{\uparrow\uparrow\uparrow\uparrow}-\Gamma^{\uparrow\uparrow\downarrow\downarrow}$
and $\Gamma^{x}=\Gamma^{y}=\frac{1}{2}(\Gamma^{\uparrow\downarrow\downarrow\uparrow}+\Gamma^{\downarrow\uparrow\uparrow\downarrow})=\Gamma^{\uparrow\downarrow\downarrow\uparrow}$.

In a local approximation, $\Sigma_k\equiv\Sigma_\nu$ and $\Gamma^{\alpha}_{kk'q}\equiv\gamma^{\alpha}_{\nu\nu'\omega}$, such as DMFT, all momentum dependence drops out of the Ward identities~\eqref{eq:Ward} and we obtain~\footnote{In an earlier publication by Hettler et al. the right-hand-side of Eq.~\eqref{eq:dmft_definition} was expressed in terms of the \textit{reducible} vertex function~\cite{Hettler00}.
In that form one does not straightforwardly realize the momentum-independence of Eq.~\eqref{eq:dmft_definition} which misled the authors to believe that DMFT does not satisfy the Ward identities.}
\begin{align}
  \Sigma_{\nu+\omega}-\Sigma_{\nu}=-\sum_{\nu'}\gamma^{\alpha}_{\nu\nu'\omega}[G_{\text{loc},\nu'+\omega}-G_{\text{loc},\nu'}].\label{eq:dmft_definition}
\end{align}
An analogous Ward identity holds for the AIM (see Appendix~\ref{app:Wardlocal}),
\begin{align}
\Sigma_{\nu+\omega}-\Sigma_\nu = -\sum_{\nu'}\gamma_{\nu\nu'\omega}^{\alpha}[g_{\nu'+\omega}-g_{\nu'}],
\label{eq:Wardimp}
\end{align}
where $\Sigma_\nu, g_\nu$ and $\gamma^\alpha_{\nu\nu'\omega}$ are the self-energy, Green's function and the irreducible vertex of the AIM, respectively. 
Hence the DMFT approximation is apparently conserving when the self-consistency condition~\eqref{eq:DMFTsc} holds.
Remarkably, DMFT arises when we attempt to construct a locally conserving approximation based on the AIM~\eqref{eq:AIM}.

Let us consider further properties of the DMFT approximation. To this  end, we introduce the (connected) susceptibilities
\begin{align}
    X^\alpha_q = -\langle \bar{\rho}^\alpha_{-q}\bar{\rho}^\alpha_q\rangle= 2\sum_{kk'}X_{kk'q}^{\alpha},
\end{align}
which are defined in terms of density fluctuations, $\bar{\rho}^\alpha(\tau)=\rho^\alpha(\tau)-\langle\rho^\alpha\rangle$.
Their local parts are given by $X^\alpha_\text{loc} = \sum_{\qv}  X^\alpha_q$.
The generalized susceptibility $X^\alpha_{kk'q}$ is related to the irreducible vertex function via the integral equation
\begin{align}
    X_{kk'q}^{\alpha}=G_{k}G_{k+q}\left[\beta \delta_{kk'}-\sum_{k''}\Gamma_{kk''q}^{\alpha}X_{k''k'q}^\alpha\right].
\label{eq:lthroughgamma}
\end{align}
Now consider the kinetic energy of the lattice. It is expressed through single-particle quantities as $E^{\text{lat}}_\text{kin}=\sum_{\kv\sigma}\varepsilon_{\kv}\langle n_{\kv\sigma}\rangle$.
In Appendix~\ref{app:asymptotefromward} we establish a relation that expresses the kinetic energy in terms of a two-particle quantity, more precisely the high-frequency behavior of the local susceptibility. The relation follows directly from the Ward identities, Eq.~\eqref{eq:Ward}:
\begin{align}
    \lim\limits_{\omega\rightarrow\infty}(\imath\omega)^2 X^{\alpha}_{\text{loc},\omega} &=-2E^{\text{lat}}_\text{kin}\label{eq:xloc_asymptote}.
\end{align}
As the Ward identities themselves, this relation connects single- and two-particle quantities.
The local impurity Ward identities~\eqref{eq:Wardimp} imply an analogous relation (see Appendix~\ref{app:WardAIM}),
\begin{align}
  \lim\limits_{\omega\rightarrow\infty}(\imath\omega)^2\chi^{\alpha}_\omega &=-2E^{\text{imp}}_\text{kin},\label{eq:imp_asymptote}
\end{align}
where $\chi^\alpha_\omega=-\langle\bar{\rho}^\alpha_{-\omega}\bar{\rho}^\alpha_\omega\rangle_{\text{imp}}$ is the impurity susceptibility and the kinetic energy of the impurity model is given by~\cite{Haule07}
\begin{equation}
E^{\text{imp}}_\text{kin}=2\sum_{\nu}\Delta_{\nu}g_{\nu}.
\label{eq:Ekinimp}
\end{equation}

DMFT is not two-particle self-consistent. As a consequence, the impurity and local lattice susceptibility differ in general.
Remarkably, however, their asymptotes are the same.
Decomposing the susceptibility into a contribution from the impurity susceptibility and a momentum-dependent correction~\cite{Hafermann14-2},
$X^{\text{DMFT}}_{\text{loc}}=\chi+X'_{\text{loc}}$, one can show that $X'_{\text{loc}}$ decays at least with $\omega^{-4}$.
Therefore, $\lim\limits_{\omega\rightarrow\infty}(\imath\omega)^2 X_{\text{loc},\omega} = \lim\limits_{\omega\rightarrow\infty}(\imath\omega)^2 \chi_\omega$.
We demonstrate this numerically in the left panel of Fig.~\ref{fig:dmft:asymptote} in the section on numerical results.
As a consequence, $E^{\text{lat}}_\text{kin}=E^{\text{imp}}_\text{kin}$ and the kinetic energy can be determined from the impurity model in DMFT.

Next, we consider the potential energy $E_{\text{pot}}=Ud_{\text{lat}}$ where $d_{\text{lat}}=\av{n_{\up} n_{\dn}}$ is the double occupancy of the lattice.
As a two-particle correlation function, $d$ is naturally computed from two-particle quantities. We denote this by a superscript '2P'. We have the following relations:
\begin{align}
      X^0_{\text{loc},\tau=0}&=-\av{\bar{\rho}^0\bar{\rho}^0}=-(\av{n}+2d^{\text{2P}}_{\text{lat}}-\av{n}^2),\label{eq:docc2P}\\
      X^z_{\text{loc},\tau=0}&=-\av{\bar{\rho}^z\bar{\rho}^z}=-(\av{n}-2d^{\text{2P}}_{\text{lat}}),
  \end{align}
where $\rho^0=n=n_\up+n_\dn, \rho^z=m=n_\up-n_\dn$ and $\av{m}=0$. Hence the double occupancy can be expressed in terms of the susceptibilities as
\begin{align}
  d^{\text{2P}}_\text{lat} &= -\frac{1}{4}\left[X^0_{\text{loc},\tau=0}-X^z_{\text{loc},\tau=0}-\av{n}^2_\text{lat}\right].
\label{eq:dlat2}
\end{align}
Similarly, $d$ may be obtained from the impurity as
\begin{align}
  d^{\text{2P}}_\text{imp} &= -\frac{1}{4}\left[\chi^0_{\tau=0}-\chi^z_{\tau=0}-\av{n}^2_\text{imp}\right]\label{eq:dimp2}.
\end{align}
By virtue of the single-particle self-consistency condition~\eqref{eq:DMFTsc} we have $\av{n}_{\text{lat}}=\av{n}_{\text{imp}}$.
Due to the missing two-particle self-consistency in DMFT however, the susceptibilities differ and we have in general $d^{\text{2P}}_\text{lat}\neq d^{\text{2P}}_\text{imp}$~\cite{vanLoon16}.
On the other hand, we can compute the double occupancies from single-particle quantities via the Migdal-Galitskii formula of the Hubbard model~\cite{Galitskii58},
\begin{align}
 d^{\text{1P}}_\text{lat} &= \frac{1}{U}\sum_{\kv\nu}G_{\kv\nu} \Sigma_{\kv\nu},\label{eq:dlat1}
\end{align}
and its counterpart of the Anderson impurity model,
\begin{align}
  d^{\text{1P}}_\text{imp}=\frac{1}{U}\sum_\nu g_\nu \Sigma_\nu\label{eq:dimp1}.
\end{align}
Making use of the single-particle self-consistency condition~\eqref{eq:DMFTsc} and of the locality of the self-energy, $\Sigma_{\kv\nu}=\Sigma_\nu$, we see that lattice and impurity double occupancies computed in this way are the same.

In summary, DMFT arises when one attempts to construct a conserving, single-particle self-consistent approximation based on the AIM.
The kinetic energy of the lattice model is equal to the kinetic energy of the impurity model.
It can be obtained from the asymptote of the local lattice susceptibility, a general feature of conserving approximations,
while in DMFT, it may also be obtained from the impurity susceptibility.
An ambiguity arises in the calculation of the double occupancy from single- and two-particle quantities: $d_\text{imp}=d^{\text{1P}}_\text{lat}\neq d^{\text{2P}}_\text{lat}$ in DMFT as a consequence of the lack of two-particle self-consistency.

We speak of several thermodynamically consistent~\cite{Aichhorn06,Janis17} ways to obtain a quantity if these yield one and the same result.
That different ways of calculating a quantity yield the same result is in general only true for an exact solution.
The kinetic energy and the $f$-sum rule (see, e.g.,~\cite{Vilk97}, see Appendix~\ref{app:fsumrule_derivation}) are examples where thermodynamic consistency between one- and two-particle level is ensured through the Ward identities~\eqref{eq:Ward}. 
Obviously, the Ward identities are insufficient for consistency in other cases, as we have seen for the double occupancy, whose value is ambiguous in DMFT.
Another important example is the inconsistency of the Schwinger-Dyson equation with the Ward identities when the reducible vertex is computed from the irreducible one through the Bethe-Salpeter equation~\cite{Janis17}.
The recently proposed QUADRILEX approach has been reported to be free of this inconsistency~\cite{Ayral16}.

We will examine in the following section to what extent the deficiencies of DMFT can be cured by two-particle self-consistency.

\section{Two-particle self-consistency}
\label{sec:2psc}

Two-particle self-consistent approximations based on an impurity model go back to extended dynamical mean-field theory (EDMFT) and its precursors~\cite{Sengupta95,Otsuki13-2,sachdev93,Si96,Kajueter96,Smith00,Chitra01}.
In these approximations, a frequency-dependent interaction is introduced in the impurity model, and its values are fixed through a self-consistency condition on a two-particle (bosonic) correlation function such as the susceptibility.
In general, we can augment the AIM of~\eqref{eq:AIM} by a dynamical interaction in all four (one charge and three spin) channels as follows:
\begin{align}
S_{\text{BFK}}=& S_{\text{AIM}}+
    \frac{1}{2}\sum_{\alpha\omega}\bar{\rho}^\alpha_{-\omega} \Lambda^\alpha_\omega\bar{\rho}^\alpha_\omega.\label{BFK}
\end{align}
We refer to this model as the Bose-Fermi-Kondo impurity (BFK) model.
$\Lambda^\alpha_{\omega}$ is a dynamical interaction which can be viewed as a bosonic bath or hybridization. We consider approximations to Green's function $G$ and to the susceptibility $X$ that are locally conserving and two-particle self-consistent. In analogy to the single-particle self-consistency condition~\eqref{eq:DMFTsc}, the retarded interactions in~\eqref{BFK} are determined through the following  condition~\cite{vanLoon16-2,Stepanov16},
\begin{align}
\chi^{\alpha}_{\omega} = X_{\text{loc},\omega}^{\alpha}.
\label{eq:2psc}
\end{align}

This self-consistency condition provides a bounded double occupancy by construction [cf. Eqs.~\eqref{eq:dlat2} and~\eqref{eq:dimp2}],
which is sufficient to suppress magnetic phase transitions in two dimensions, as shown by Vilk and Tremblay~\cite{Vilk97}.

The Ward-identities~\eqref{eq:Ward} relate single-particle quantities (Green's function and self-energies) to two-particle quantities (vertex functions and susceptibilities).
We study the interplay between the requirement of local conservation and the self-consistency conditions~\eqref{eq:DMFTsc} and~\eqref{eq:2psc}.

We differentiate between two kinds of approaches: (i) The case of Ising-type ($S_{z}$) coupling is characterized by a finite $\Lambda^z_\omega$ and $\Lambda^{0}_\omega$, while we set $\Lambda^{x,y}_\omega = 0$. That is, we require self-consistency~\eqref{eq:2psc} only in the charge- and one of the spin channels, i.e., for $\alpha=0,z$.
(ii) For rotationally invariant Heisenberg-type coupling, all retarded interactions $\Lambda_{\omega}^{\alpha}$ for $\alpha=0,x,y,z$ are finite and determined by~\eqref{eq:2psc}.
The retarded interactions cause shifts in the Hubbard interaction and the chemical potential which are discussed in Appendix~\ref{app:impurity:hamiltonian}.

\subsection{Ising-type coupling}
\label{section:sz-coupling}

In the discussion of the kinetic energy in the context of DMFT we have shown that, assuming local conservation, it can be expressed in terms of the asymptotic behavior of the susceptibility.
For the lattice we have by virtue of local conservation $\lim\limits_{\omega\rightarrow\infty}(\imath\omega)^2 X^{0,z}_{\text{loc},\omega} =-2E_\text{kin}^{\text{lat}}$, while on the impurity,
$\lim\limits_{\omega\rightarrow\infty}(\imath\omega)^2\chi^{0,z}_\omega =-2E_{\text{kin}}^{\text{imp}}$ holds (cf. Appendix~\ref{app:asymptotics:impurity}). 
In this case, the kinetic energies computed in the two ways are equal by means of the two-particle self-consistency~\eqref{eq:2psc}.
The kinetic energy can therefore be obtained from the impurity model, as in DMFT. Note that $E_\text{kin}^{\text{lat}}=E_{\text{kin}}^{\text{imp}}$ may be determined from the charge or spin susceptibility alike.

We saw previously that the double occupancy computed from two-particle quantities, $d_{\text{lat}}^{\text{2P}}$ and $d_{\text{imp}}^{\text{2P}}$, can be expressed in terms of the local susceptibilities $X_{\text{loc}}^{0,z}$ and $\chi^{0,z}$, respectively [Eqs.~\eqref{eq:dlat2} and~\eqref{eq:dimp2}].
While these differ in DMFT in general, the two-particle self-consistency ensures that $d_{\text{lat}}^{\text{2P}}=d_{\text{imp}}^{\text{2P}}$.
Using single-particle quantities, we can still compute it on the lattice using the Migdal-Galitskii formula, Eq.~\eqref{eq:dlat1}.
In a local approximation to the self-energy and with the single-particle self-consistency condition~\eqref{eq:DMFTsc}, we can express the double occupancy in terms of the impurity self-energy and Green's function, $d^{\text{1P}}_\text{lat} = (1/U)\sum_\nu g_\nu \Sigma_\nu$.
In contrast to DMFT however, this expression is \emph{not} equal to the impurity double occupancy. 
Because the Migdal-Galitskii formula involves the potential energy, it is comprehensible that the retarded interactions will affect it. In Appendix~\ref{app:migdal} we derive the double occupancy for the Bose-Fermi-Kondo model, with the result
\begin{align}
 d^{\text{1P}}_\text{imp} &= \frac{1}{2\tilde{U}}\left[2\sum_\nu g_\nu \Sigma_\nu+\sum_{\omega,\alpha}\tilde{\Lambda}^\alpha_\omega(\chi^\alpha_\omega-\beta\av{\rho^\alpha}^2\delta_\omega)\right].
    \label{eq:dimp1_bfk}
\end{align}
Here $\tilde{U}=U+\Lambda^0_\infty-\Lambda^z_\infty$ contains the asymptotic part of the retarded interaction $\Lambda^{\alpha}_\omega=\Lambda^{\alpha}_\infty+\tilde{\Lambda}^{\alpha}_\omega$ (cf. Appendix~\ref{app:impurity:hamiltonian}). The summation in the second term in brackets in Eq.~\eqref{eq:dimp1_bfk} in general runs over all channels, $\alpha=0,x,y,z$. In the case of Ising-type coupling, only the retarded interaction in the two channels $\alpha=0,z$ is non-zero.

Because we assume that we solve the impurity model exactly, we have $d^{\text{1P}}_\text{imp}=d^{\text{2P}}_\text{imp}\equiv d_\text{imp}$ (so that we can drop the superscript indices). While $d^{\text{2P}}_\text{lat}=d_\text{imp}$, the second term in~\eqref{eq:dimp1_bfk} will in general lead to $d^{\text{1P}}_\text{lat}\neq d_\text{imp}$. The double occupancy computed from two-particle quantities $d^{\text{2P}}$ is consistent with that of the impurity (because of two-particle self-consistency), while that obtained from single-particle quantities is not. 
This situation is exactly opposite to DMFT, where $d^{\text{1P}}$ is consistent. We see that while the retarded interactions allow us to enforce consistency of $d^{\text{2P}}$, they simultaneously undermine the consistency of $d^{\text{1P}}$.
We demonstrate this numerically in Fig.~\ref{fig:docc} for the two-particle self-consistent approximation presented in Sec.~\ref{section:scdb} representative of Ising-type coupling and compare to DMFT.

\subsection{Heisenberg-type coupling}
\label{section:3d-coupling}

In the case of a Heisenberg-type coupling, all retarded interactions $\Lambda_{\omega}^{\alpha}$ are in general non-zero. We fix their values through the self-consistency condition~\eqref{eq:2psc}, as before, and consider the $SU(2)$-symmetric case with $\Lambda^x=\Lambda^y=\Lambda^z$.
We further assume that the Ward identities hold. As a consequence, the relation (cf. Appendix~\ref{app:asymptotefromward})
\begin{align}
  \lim\limits_{\omega\rightarrow\infty}(\imath\omega)^2 X^{\alpha}_{\text{loc},\omega} =-2E_\text{kin}\label{eq:xloc_asymptote_3d}
\end{align}
holds in all channels $\alpha=0,x,y,z$. 

On the impurity model, contrary to the case of Ising-type coupling, we now have separate relations for the charge and spin susceptibilities (Appendix~\ref{app:asymptotics:impurity}):
\begin{align}
  \lim\limits_{\omega\rightarrow\infty}(\imath\omega)^2\chi^0_\omega &=-4\sum_\nu\Delta_\nu g_\nu,\label{eq:imp_asymptote_3dch}\\
  \lim\limits_{\omega\rightarrow\infty}(\imath\omega)^2\chi^z_\omega &=-4\sum_\nu\Delta_\nu g_\nu+4\sum_{\omega',\alpha=x,y}\tilde{\Lambda}^\alpha_{\omega'}\chi^\alpha_{\omega'}.\label{eq:imp_asymptote_3dsp}
\end{align}
The corresponding relations for $\chi^{x,y}$ are obtained by permuting $x,y,z$ in~\eqref{eq:imp_asymptote_3dsp}.
We see that in presence of retarded spin interactions the asymptote of the impurity spin susceptibility in Eq.~\eqref{eq:imp_asymptote_3dsp} no longer equals the kinetic energy.
In addition, the asymptotes of the charge and spin channels are different.
By virtue of the self-consistency condition, $\chi_{\omega}^{\alpha}=X_{\text{loc},\omega}^{\alpha}$, this must also hold for the asymptote of $X_{\text{loc},\omega}^{\alpha}$.
Eq.~\eqref{eq:xloc_asymptote_3d}, on the other hand, implies that the asymptote of the local susceptibility must be equal in all channels. We therefore conclude that there is no two-particle self-consistent approximation employing the self-consistency condition~\eqref{eq:2psc}, which at the same time is locally conserving
~\footnote{This only holds if $\sum_{\omega',\alpha=x,y}\tilde{\Lambda}^\alpha_{\omega'}\chi^\alpha_{\omega'}$ is nonzero. The term does not vanish in our calculations and can only do so for an unphysical $\tilde{\Lambda}^\alpha_{\omega}$ which changes its sign for different $\omega$ [as $\chi$ does not, cf. Eq.~\eqref{discretelambda}].}.
We note that in the case of Heisenberg-type coupling, the conclusions regarding the potential energy remain the same as in the Ising-type coupling. In particular the equation~\eqref{eq:dimp1_bfk} still holds.
We provide a numerical example of Eqs.~\eqref{eq:imp_asymptote_3dch} and~\eqref{eq:imp_asymptote_3dsp} in the right panel of Fig.~\ref{fig:doped:chi:imp:asymptote} in Appendix~\ref{app:asymptotics:impurity}.

\section{Ward identities and retarded spin-spin interactions}
\label{section:conservation_scdb}

In this section we show that the Ward identities of the Bose-Fermi-Kondo model are incompatible with the Ward identities~\eqref{eq:Ward} of the Hubbard model.
We identify this as the root cause of our earlier finding in the previous section, that no locally conserving approximation can be obtained in case of a Heisenberg-type coupling.

In the Hamiltonian formulation of the Bose-Fermi-Kondo model (cf. Appendix~\ref{bose_anderson}) the retarded interactions enter as density-boson couplings $\propto \phi^\alpha \rho^\alpha$.
Here, $\phi=b^\dagger+b$ are bosonic operators which commute with all fermions. Integrating out the bosons in the functional integral yields the effective impurity action~\eqref{BFK}.

Since the Ward identities are Green's function equivalents of the continuity equations, they describe conservation of the charge- and spin-currents.
These currents may be caused by kinetic and interaction contributions.

\begin{figure}[t]
 \includegraphics[]{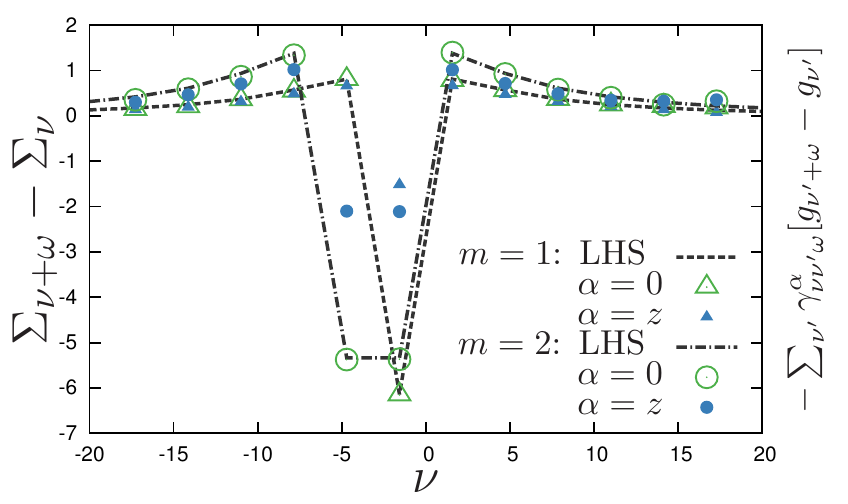}
 \caption{(Color online) Test of Eq.~\eqref{eq:Wardimp} for an isotropic retarded spin-spin interaction $\tilde{\Lambda}^x=\tilde{\Lambda}^y=\tilde{\Lambda}^z$ and a retarded charge-charge interaction $\tilde{\Lambda}^0$.
 The imaginary part of the left-hand-side (dashed black lines) and of the right-hand-side (symbols) of Eq.~\eqref{eq:Wardimp} is drawn at the first two bosonic Matsubara frequencies $\omega_{m=1,2}$.
 Eq.~\eqref{eq:Wardimp} holds in the charge channel (open green symbols) but is violated in the spin channels (filled blue symbols).
 This test was performed at $\beta=2$ and $U=6$ with a conducting bath $\Delta$.
 The violation of Eq.~\eqref{eq:Wardimp} in the spin channels depends on the magnitude of $\tilde{\Lambda}^{x,y,z}$, which was chosen large for demonstration purposes.
 The data shown in this figure was produced with the CTQMC solver presented in reference~\cite{Otsuki13}.}
 \label{fig:ward3d}
\end{figure}

Regarding the latter, we notice two properties: (i) None of the retarded interactions contribute to the charge-current, that is $[\rho^{0},\phi^\alpha \rho^\alpha]=\phi^\alpha [n,\rho^\alpha]=0$ for $\alpha=0,x,y,z$.
(ii) The spin-current on the other hand has contributions from the retarded spin-spin interactions $\Lambda^\beta$ due to non-commutativity of the spin operators,
$[\rho^{\alpha},\phi^\beta \rho^\beta]=2\imath\phi^\beta\sum_\gamma\varepsilon_{\alpha\beta\gamma}\rho^\gamma$ for $\alpha,\beta=x,y,z$.
We show in Appendix~\ref{app:Wardlocal} that the resulting Ward identities contain an additional term that couples the retarded spin interaction to a three-particle correlation function~\cite{rostamiTBP}.
As a consequence, they cannot be brought into the form of the local Ward identities~\eqref{eq:Wardimp}.
We emphasize that this does not imply a violation of spin conservation in the Bose-Fermi-Kondo model.
The issue is instead that the Ward identities accounting for spin conservation simply have a form different from Eq.~\eqref{eq:Wardimp}.
It therefore seems plausible that conservation on the level of the BFK model does not imply that the local Ward identities~\eqref{eq:Wardimp} are fulfilled.
That they are indeed violated in general is illustrated numerically in Fig.~\ref{fig:ward3d} by plotting the left- and right-hand sides of Eq.~\eqref{eq:Wardimp} for finite $\tilde{\Lambda}^{x,y,z}$.

In order to understand the consequences for constructing conserving approximations based on an impurity model, we recall that in a local approximation to the self-energy and irreducible vertex function the local Ward identities~\eqref{eq:dmft_definition} are sufficient to guarantee that the approximation is conserving.
In the case of DMFT, with the self-consistency condition $G_{\text{loc}}=g$, they \emph{coincide} with the Ward identities of the AIM, so that DMFT is conserving. In presence of retarded spin-spin interactions this is no longer the case and, as we have seen numerically, this equation in general is violated. This can be seen as follows: Eq.~\eqref{eq:Wardimp} implies that the tails of the local susceptibilities must be identical independent of the channel index $\alpha$. We show this in Appendix~\ref{app:asymptotefromward} and~\ref{app:WardAIM}. In the previous section, we have seen however, that for the Heisenberg-type coupling they are different because of the retarded interaction [cf. Eqs.~\eqref{eq:imp_asymptote_3dch} and~\eqref{eq:imp_asymptote_3dsp}]. \eqref{eq:Wardimp} must therefore be violated and the approximation is not conserving.

\begin{figure}[t]
 \includegraphics[]{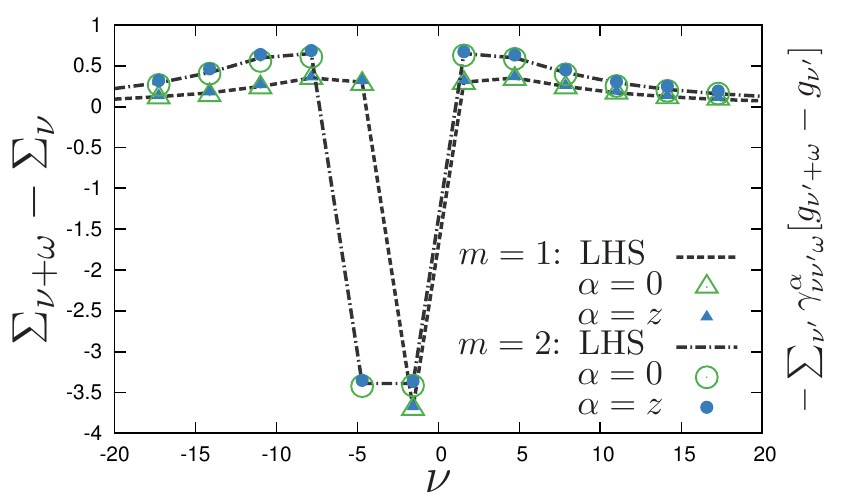}
 \caption{(Color online) Test of Eq.~\eqref{eq:Wardimp} for a retarded spin-spin interaction $\tilde{\Lambda}^z$ in the $z$-channel
     and a retarded charge-charge interaction $\tilde{\Lambda}^0$. Eq.~\eqref{eq:Wardimp} holds in the channels $\alpha=0,z$.
 Parameters as in Fig.~\ref{fig:ward3d}, except $\tilde{\Lambda}^x=\tilde{\Lambda}^y=0$.}
 \label{fig:ward}
\end{figure}

In the case of Ising-type coupling, the retarded interaction $\Lambda^z$ in the longitudinal spin channel contributes to the currents in the transversal spin channels of the impurity.
The violation of the local Ward identites~\eqref{eq:Wardimp} thus affects only the transversal spin channels, while the longitudinal spin channel itself remains unaffected.
That the Ward identity in the longitudinal spin channel indeed holds under these circumstances is demonstrated in Fig.~\ref{fig:ward}.

\section{An example: Two-particle self-consistent DMFT}
\label{section:scdb}

We have discussed the general conditions for a conserving and two-particle self-consistent approximation in Sec.~\ref{sec:2psc}. Here we construct a concrete example.
As we have seen, an approximation that satisfies the two particle self-consistency condition~\eqref{eq:2psc} can be conserving in the charge- and at most one of the spin channels. One may refer to this approximation as two-particle self-consistent DMFT.

We compute the lattice susceptibility in this approach according to
  \begin{align}
      X^{\alpha}_{\qv\omega}=\left[\left(X^{\text{DMFT},\alpha}_{\qv\omega}\right)^{-1}+\Lambda^{\alpha}_\omega\right]^{-1}.\label{eq:xdb}
  \end{align}
The particular form~\eqref{eq:xdb} of the susceptibility can be motivated in the DB approach~\cite{Rubtsov12}. In this form the retarded interaction is reminiscent of the Moriya~$\Lambda$ correction employed in D$\Gamma$A.
Here $\Lambda$ however depends on frequency, while in D$\Gamma$A it is instantaneous.
We emphasize that the way of calculating the susceptibility and its particular form do not change the conserving character of the theory (see Sec.~\ref{section:sz-coupling} and results in Sec.~\ref{sec:results} below), but will of course affect the results. 

\begin{figure}[t]
 \includegraphics[]{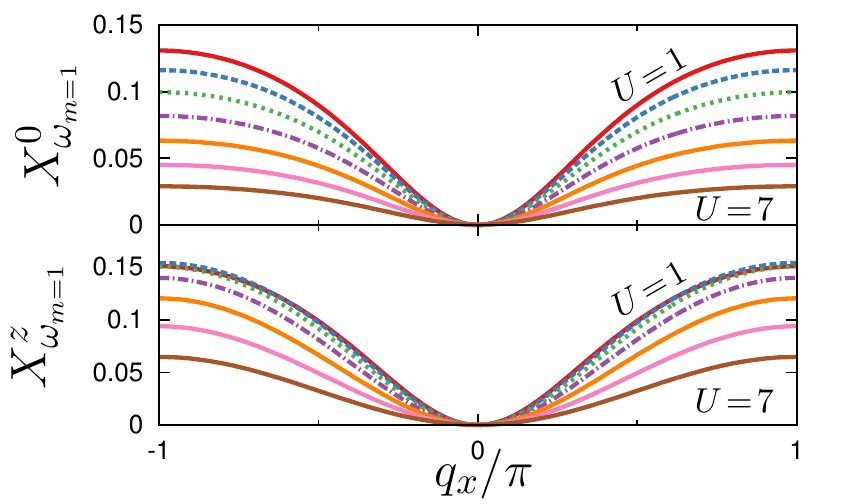}
 \caption{(Color online) Susceptibility at the first Matsubara frequency in the two-particle self-consistent DMFT. Shown is a momentum cross-section at $q_y=0$ for different values of $U$ ($\beta=2$).
     The vanishing of the susceptibility for $\qv\rightarrow 0$ with $X\sim |\qv|^2$ is a necessary condition for global conservation, see text.}
 \label{fig:x_wn1}
\end{figure}

In the above, $X^{\text{DMFT},\alpha}$ denotes the susceptibility computed as in DMFT in the standard way~\cite{Georges96}, including vertex corrections.
This amounts to approximating the irreducible vertex function of the lattice with its local counterpart on the impurity, $\Gamma^{\alpha}_{kk'q}\equiv\gamma^{\alpha}_{\nu\nu'\omega}$, in the channels $\alpha=0,z$.
We compute the generalized susceptibility from the integral equation $X_{kk'q}^{\alpha}=G_{k}G_{k+q}\left[\beta\delta_{kk'}-\sum_{k''}\gamma_{\nu\nu''\omega}^{\alpha}X_{k''k'q}^\alpha\right]$. The susceptibilities are obtained from the latter by tracing out $k,k'$:  $X^{\text{DMFT},\alpha}_q=2\sum_{kk'}X_{kk'q}^{\alpha}$.
We emphasize that the label 'DMFT' merely indicates that $X^{\text{DMFT},\alpha}$ is computed as in DMFT. Its value will differ from the DMFT susceptibility, because the impurity model is different.

The BFK model can be solved accurately using a suitably generalized continuous-time quantum Monte Carlo (CTQMC) algorithm. In weak-coupling CTQMC, the inclusion of these terms is straightforward~\cite{Rubtsov05}. In strong-coupling CTQMC the impurity model can be solved in the segment representation when only $\Lambda^{0}$ or $\Lambda^{z}$ are included~\cite{Werner07,Werner10,Hafermann13}. For the general case of a vector bosonic field (not considered in the numerical results of this section), the algorithm simultaneously performs a hybridization expansion and an interaction expansion with respect to the spin-off-diagonal interactions $\Lambda^{x,y}$~\cite{Otsuki13}. 
Here we compute the correlation functions $g_\nu$ and $\chi^\alpha_\omega$, the self-energy $\Sigma_\nu$ and the irreducible vertex function $\gamma_{\nu\nu'\omega}^{\alpha}$ for $\alpha=0,z$ using a strong coupling quantum Monte Carlo solver~\cite{Hafermann13} with improved estimators adapted to treat the retarded interactions~\cite{Hafermann12,Hafermann14}.

\begin{figure}[t]
    \centering
  \begin{minipage}[b]{0.235\textwidth}
      \includegraphics[width=\textwidth]{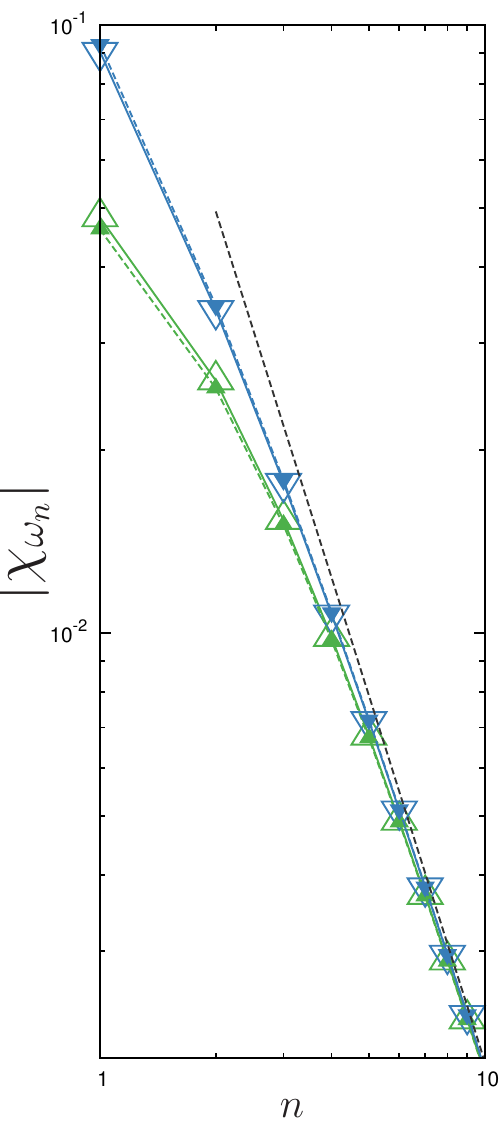}
  \end{minipage}
\hfill
  \begin{minipage}[b]{0.235\textwidth}
      \includegraphics[width=\textwidth]{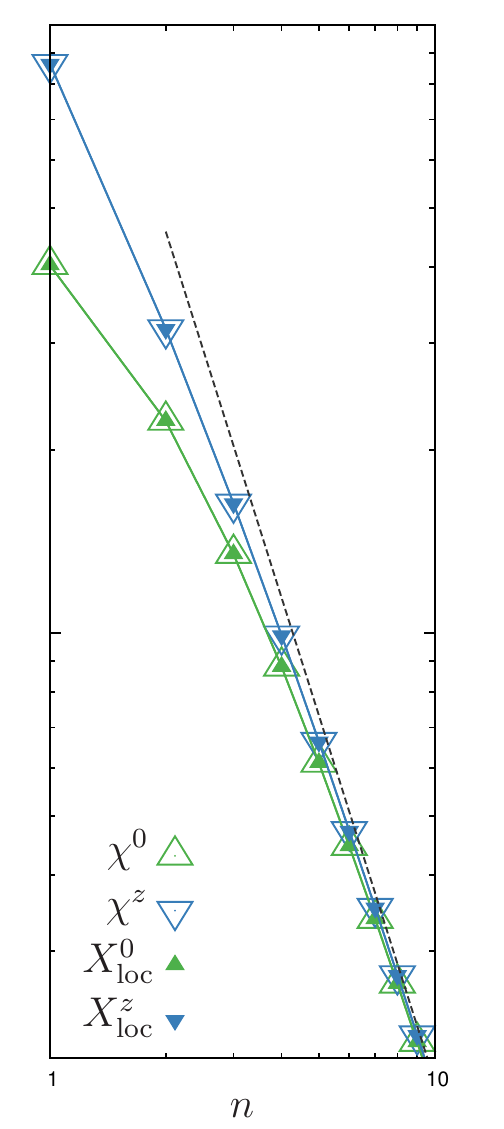}
  \end{minipage}
 \caption{\label{fig:dmft:asymptote} (Color online) High frequency behavior of $\chi$ (open triangles, bold lines) and $X_{\text{loc}}$ (filled triangles, dashed lines) in DMFT (left) and in the two-particle self-consistent DMFT (right).
   The dashed black lines indicate the asymptotes $-2E_\text{kin}/\omega^2$ computed from~\eqref{eq:imp_asymptote}.
   The charge (green) and spin (blue) susceptibility approach the same asymptote in both approximations.
  }
\end{figure}

The calculation procedure is as follows: We start from initial values for the hybridization $\Delta_{\nu}$ and retarded interactions $\Lambda^{\alpha}_{\omega}$, which specify the BFK impurity model~\eqref{BFK}. After solving the model, we evaluate the lattice susceptibility~\eqref{eq:xdb}. The Green's function is computed from the impurity self-energy in the same way as in DMFT:
\begin{align}
G_{\kv\nu}^{-1} = i\nu+\mu-\epsilon_{\kv}-\Sigma_{\nu}^{\text{imp}}.
\end{align}
The local parts of $G_{\kv\nu}$ and $X^{\alpha}_{\qv\omega}$ will in general be different from the impurity quantities $g_\nu$ and $\chi^{\alpha}_{\omega}$. We update the hybridization $\Delta_{\nu}$ and retarded interactions $\Lambda^{0}_{\omega}$, $\Lambda^{z}_{\omega}$ simultaneously and iteratively, until the conditions $G_{\text{loc},\nu}=g_\nu$ and $X^\alpha_{\text{loc},\omega}=\chi^\alpha_\omega$ for $\alpha=0,z$ are satisfied.

\subsection{Numerical results}
\label{sec:results}

Let us now turn to the discussion of numerical results of the two-particle self-consistent DMFT.
In the following we use parameters $U=6$, $T=0.5$ (in units of $t$), which is somewhat above the DMFT N\'eel temperature $T\approx 0.35$.

In Fig.~\ref{fig:ward} we illustrate numerically that contrary to Heisenberg-type coupling (cf. Fig.~\ref{fig:ward3d}) the local Ward identities~\eqref{eq:Wardimp} hold in the considered channels, $\alpha=0,z$.
As shown in Appendix~\ref{app:asymptotefromward}, this implies $X^{\text{DMFT},\alpha}_{\qv=0,\omega\neq 0}=0$. Inserting this into Eq.~\eqref{eq:xdb} it follows that $X^\alpha_{\qv=0,\omega\neq 0}=0$, which is a necessary condition for conservation of the total density, i.e., $\omega\rho_{\qv=0,\omega}=0$.
Thus, $X^0$ and $X^z$ are at least globally conserving.

\begin{figure}
    \includegraphics[width=0.49\textwidth]{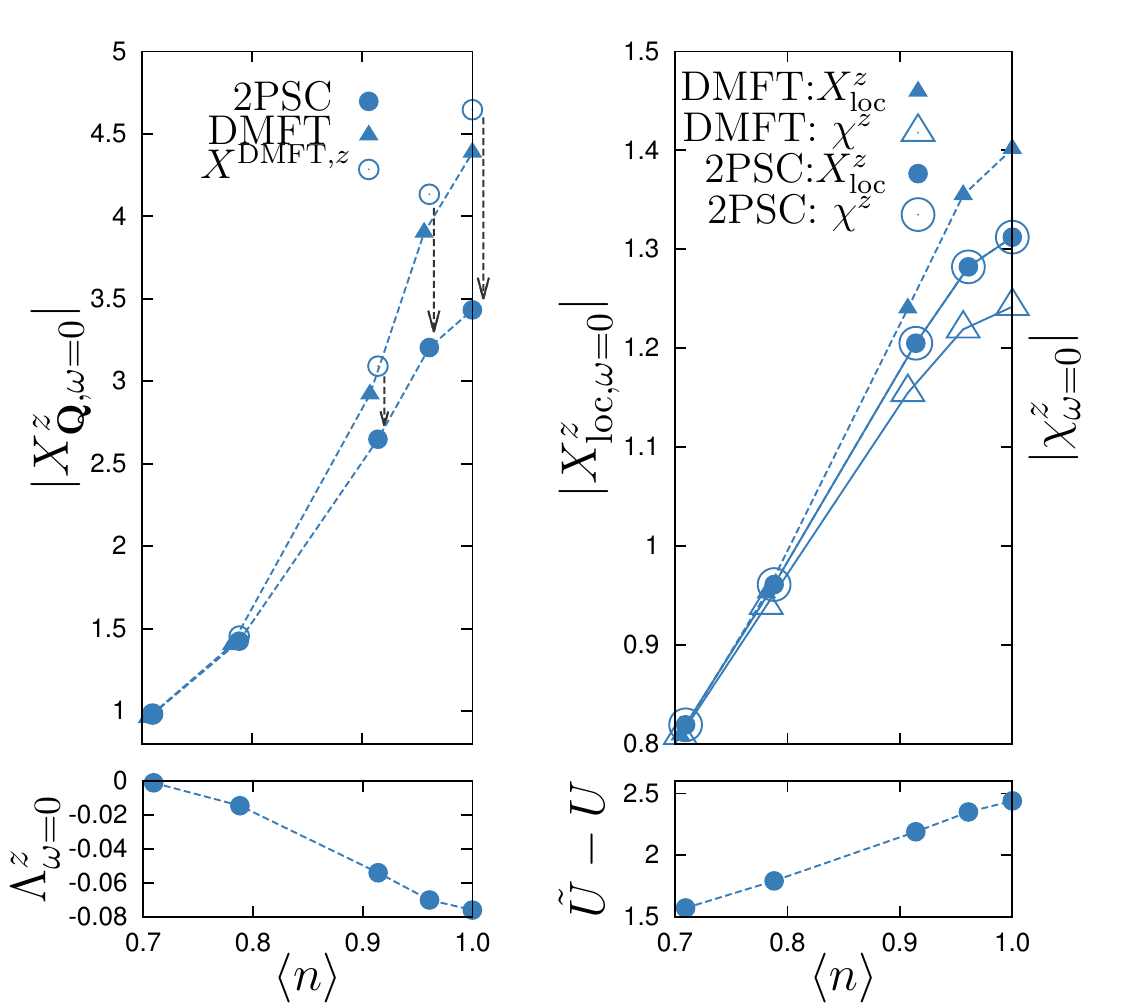}
    \caption{Top: Static susceptibilities of the Hubbard model~\eqref{hubbardmodel} at $U=6, \beta=2$
    in DMFT (triangles) and in the 2P self-consistent approximation~\eqref{eq:xdb} (circles) as a function of the density $\langle n\rangle$.
    Left: Lattice susceptibility $X^z_{\mathbf{Q},\omega=0}$ at $\mathbf{Q}=(\pi,\pi)$. Right: Local lattice ($X^z_{\text{loc}}$) and impurity ($\chi^z$) susceptibility.
    In DMFT (open and filled triangles), the local susceptibility $X^z_{\text{loc}}$ (full triangles, dashed lines) is larger than the impurity susceptibility $\chi^z$ (open triangles, bold lines). They coincide in the 2P self-consistent approximation (open and filled circles). Bottom, left: Static component of the retarded spin-spin interaction $\Lambda^z_{\omega=0}$. Bottom, right: Effective Hubbard repulsion $\tilde{U}$.
    }
\label{fig:doped:X:sz}
\end{figure}

Fig.~\ref{fig:x_wn1} illustrates that $X^0$ and $X^z$ indeed vanish in the limit $|\qv|\to 0$ for finite frequencies ($\omega_{m=1}$ in this case).
We note that due to the retarded spin interaction $\tilde{\Lambda}^z$ this approximation is not conserving in the $x$ and $y$ channels (cf. discussion in Sec.~\ref{section:conservation_scdb} and Appendix~\ref{app:Wardlocal}).

The right panel of Fig.~\ref{fig:dmft:asymptote} demonstrates the equivalence of the impurity and local lattice susceptibility in two-particle self-consistent DMFT.
In this approximation the charge and the longitudinal spin susceptibility approach the same asymptote.
This is required by the conservation laws and also satisfied in DMFT (see left panel, cf. Sec.~\ref{sec:DMFT} and~\ref{section:sz-coupling}).
 
In the top left panel of Fig.~\ref{fig:doped:X:sz} we compare the static spin susceptibility $X_{\mathbf{Q},\omega=0}$ at $\mathbf{Q}=(\pi,\pi)$ in DMFT,
two-particle self-consistent DMFT and the quantity $X_{\mathbf{Q},\omega=0}^{\text{DMFT}}$ in~\eqref{eq:xdb} at and near half-filling.
The increase of $X_{\mathbf{Q},\omega=0}^{\text{DMFT}}$ compared to its value in standard DMFT is consistent with an enhanced local interaction on the impurity model $\tilde{U}-U=\Lambda^0_\infty-\Lambda^z_\infty>0$ as seen in the bottom right panel of Fig.~\ref{fig:doped:X:sz}.
Concomitantly, we also find a larger leading eigenvalue of the Bethe-Salpeter equation in the two-particle self-consistent DMFT (not shown).
\begin{figure}[t]
      \includegraphics[width=0.49\textwidth]{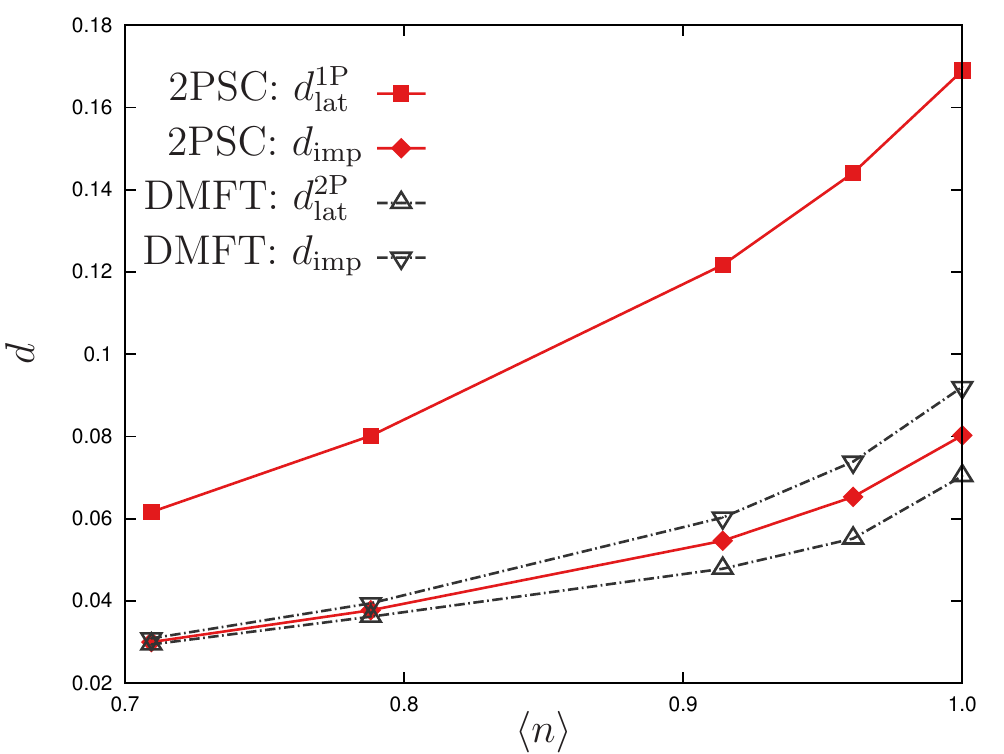}
 \caption{\label{fig:docc} (Color online) Double occupancy as a function of filling. The figure illustrates the inconsistency of $d_\text{imp}$ and $d^{\text{2P}}_\text{lat}$ in DMFT and of $d_\text{imp}$ and $d^{\text{1P}}_\text{lat}$ in 2PSC.
The impurity double occupancy computed from the susceptibilities~\eqref{eq:dimp2} and the Migdal-Galitskii formula~\eqref{eq:dimp1_bfk} yield the same result because we solve the impurity model exactly.}
\end{figure}

In the top right panel we see that the local susceptibility of the converged solution lies between the two values of DMFT.
The effect of the two-particle self-consistency is larger close to half-filling, where antiferromagnetic fluctuations are strongest.
Compared to DMFT, the increase of $\chi_{\omega=0}$ in the two-particle self-consistent method correlates with the large enhancement $U\rightarrow\tilde{U}$ of the on-site interaction in the impurity model.

Finally, we see in the top left panel of Fig.~\ref{fig:doped:X:sz} that $X^z_{\mathbf{Q},\omega=0}$ is significantly reduced compared to $X^{\text{DMFT}}$ due to $\Lambda$ which acts as a cutoff.
This reduction (marked by downward arrows) becomes larger with a larger absolute value of the cutoff $\Lambda_{\omega=0}$, as can be seen in the bottom left panel.

While the lattice susceptibility~\eqref{eq:xdb} can in principle be defined without the dynamical cutoff $\Lambda$, this solution gives results closer to benchmarks:
The double occupancy computed from the susceptibilities according to Eq.~\eqref{eq:dlat2} or~\eqref{eq:dimp2} gives a result that is closer to DCA benchmarks than either of the two values that are obtained in DMFT~\cite{vanLoon16}.
Despite two-particle self-consistency, the double occupancy is nevertheless inconsistent between one- and two-particle level, as discussed in Sec.~\ref{section:sz-coupling} and demonstrated in Fig.~\ref{fig:docc}.

We note that this approximation has several issues:
The dynamic part of the retarded interaction in the spin channel $\tilde{\Lambda}_\omega=\Lambda_\omega-\Lambda_\infty$ is positive.
This corresponds to negative energies of the bosons [cf.~\eqref{discretelambda}] and is unphysical. The impurity model can nevertheless be solved in the QMC solver in the segment picture.
Secondly, the asymptotic behavior of the self-energy is modified due to the retarded interactions~\cite{Hafermann14}. Since DMFT produces the correct asymptotic behavior~\cite{Rohringer16}, the high-frequency tail of the self-energy in this approximation is no longer exact.
Finally, even though the approximation suppresses a magnetic phase transition in two dimensions, the momentum-independent cutoff leads to an unphysical plateau of the susceptibility $X_{\qv,\omega=0}=\Lambda^{-1}_{\omega=0}$ for all momenta $\qv$ in the vicinity of $\vc{Q}$ for which $X^{\text{DMFT}}_{\qv,\omega=0}$ diverges when approaching the DMFT N\'eel temperature.

\section{Discussion}
\label{sec:discussion}

We have discussed the conservation of charge and spin in two-particle self-consistent extensions of DMFT for the Hubbard model. 
For large interaction, the Hubbard model approximately maps to the Heisenberg model and is hence dominated by spin fluctuations.
The motivation for including a retarded spin-spin interaction into the impurity model is to account for these fluctuations.

As we have seen however, introducing a retarded interaction in the longitudinal spin channel leads to a violation of conservation in the transversal spin channels of the lattice approximation. 
Moreover, a retarded interaction in all three spin channels violates conservation on the lattice in all spin channels (cf. Table~\ref{table:conservation}).

We have argued that this is related to the fact that the Ward identities of the lattice and impurity are incompatible.
To make sense of this physically, we recall that an interaction only conserves the local charge- or spin-density if it commutes with the corresponding observable.
The retarded charge-charge interaction commutes with the charge density and therefore preserves charge on the impurity.
In other words, the bosons that mediate the retarded interaction do not carry a charge.
On the other hand, the transversal components of the retarded spin-spin interaction do \emph{not} commute with the longitudinal spin density operator.
Consequently, the spin bosons carry spin: Acting with the operator $S^{+}$ or $S^{-}$ on the impurity flips the spin of an electron by one quantum, which is carried by the boson. This leads to spin currents onto and off the impurity, which manifest themselves in the impurity Ward identities.
These currents have no analogue in the Hubbard model, where the motion of spin inevitably involves motion of charge. The latter is accounted for by the fermionic hybridization function.
In essence, the introduction of the retarded spin-spin interactions in order to achieve two-particle self-consistency causes a ``spin leak'' in the lattice approximation.
One may speculate that if the interaction of a lattice model conserves a local density, the local reference system should have the same property.
\begin{table}[t]
\begin{tabular}{ l | c || c | c || c | c | c}
    & $\Lambda$ & $d^{\text{1P}}$ & $d^{\text{2P}}$ & charge & spin-$z$ & spin-$x,y$ \\\hline
      DMFT & - & \checkmark & - & \checkmark & \checkmark & \checkmark \\\hline
     Ising & $0,z$ & - & \checkmark & \checkmark & \checkmark & - \\\hline
Heisenberg & $0,x,y,z$ & - & \checkmark & \checkmark & - & - \\\hline
\end{tabular}
\caption{\label{table:conservation} A summary of the main results for DMFT and 2PSC methods, respectively.
    The second column indicates which retarded interactions $\Lambda^\alpha$ act on the impurity.
    The third and fourth column show if the double occupancy $d$ is consistent between impurity and lattice on the 1P and 2P level, respectively [see Sec.~\ref{sec:DMFT} and~\ref{sec:2psc}].
    The remaining columns list the channels in which local conservation is satisfied [see Eqs.~\eqref{eq:Ward}-\eqref{eq:Wardimp} for DMFT, and Sec.~\ref{section:3d-coupling} and~\ref{section:conservation_scdb} for 2PSC].
}
\end{table}

Even though the Hubbard model maps to a Heisenberg model for strong coupling, spin conservation is violated due to the exchange interaction on the impurity.
This is no contradiction because the Heisenberg model is an effective low-energy model.
The Ward identities however imply the equivalence of the charge- and spin susceptibilities and excitations at high energies for any finite value of $U$ [cf. Eq.~\eqref{eq:xloc_asymptote} in Sec.~\ref{sec:DMFT}]. Indeed, the effective exchange coupling $J=-4t^{2}/U$ involves two hopping processes and thus virtual high-energy charge excitations.

In the $t$-$J$ model, on the other hand, part of the spin currents is caused by the exchange interaction on the lattice. This part is decoupled from the charge current.
Contrary to the Hubbard model, a spin current on the impurity that is decoupled from the charge current a priori poses no problem and is even necessary.
However, there remains the problem of finding a two-particle self-consistency condition which satisfies the Ward identities of the $t$-$J$ model.

This brings us to a last point, namely a possible way out of this dilemma.
We recall that our conclusions about the conserving character of the considered approximations follow from the two-particle self-consistency condition $\chi^{\alpha}_{\omega} = X_{\text{loc},\omega}^{\alpha}$, which seems like a natural choice.
We can therefore not rule out the possibility that a different prescription exists, such that conservation is satisfied.
In view of the above arguments, this seems unlikely in case of the Hubbard model, but more promising for the $t$-$J$ model.

\section{Conclusions}
\label{sec:conclusions}

We have investigated the interplay between the requirement of conservation of an approximation and two-particle self-consistency. 
Retarded interactions are required to enforce two-particle self-consistency, but their presence leads to problems. While the ambiguity in computing the DMFT double occupancy from two-particle quantities is resolved, the retarded interaction instead introduces an ambiguity in the calculation of the double occupancy from single-particle quantities.

More importantly, the Ward identities of the resulting impurity model are no longer compatible with the lattice Ward identities.
As a consequence, we found that it is impossible to construct a two-particle self-consistent approximation to the Hubbard model which simultaneously fulfills the lattice Ward identities in the charge and all spin channels.

A conserving two-particle self-consistent approximation can be obtained when restricting self-consistency to the charge and one of the spin channels.
We have used this to construct a two-particle self-consistent version of DMFT, which provably obeys global conservation laws and which resolves the ambiguity in the calculation of the double occupancy on the two-particle level.
While this approximation suppresses a magnetic phase transition in two dimensions and yields results for the double occupancy which are closer to benchmarks than either of the two DMFT values,
it however has several issues which make it impractical, in particular at low temperature.
Our results imply constraints for the construction of two-particle self-consistent diagrammatic extensions.

Finally, we have seen that DMFT arises naturally when constructing a conserving approximation based on the Anderson impurity model.
It may be possible to derive the cellular DMFT from the Ward identities, avoiding the cavity construction.

\acknowledgments

We thank the referees for constructive suggestions that have led to an improvement of this work.
F.K. likes to thank G. Rohringer and A. Toschi for a useful discussion on Ward identities.
F.K. and A.L. are supported by the DFG-SFB668 program. E.G.C.P.v.L. and M.I.K. acknowledge support from ERC Advanced Grant 338957 FEMTO/NANO. 
The auxiliary impurity model was solved using a modified version of the open source CT-HYB solver~\cite{Hafermann13} based on the ALPS libraries~\cite{ALPS2}.
Computational resources were provided by the HLRN-cluster under Project No. hhp 00030.

\appendix
\section{Lattice Ward identities}
\label{app:Wardlat}
In this Appendix we detail the derivation of the Ward identities of a quantum lattice model with Hamiltonian $H=H_0+H_{\text{int}}$, where $H_0=\sum_{\kv\sigma}\varepsilon_{\kv}c^\dagger_{\kv\sigma}c_{\kv\sigma}$ is the non-interacting Hamiltonian and $H_{\text{int}}$ is the interaction part. The Ward identities for the continuum can be found in textbooks~\cite{landau13,Schrieffer99}. Derivations for quantum lattice systems have been given in, e.g.,~\cite{Vollhardt80,Revzen89,Dare94,Toyoda97,Janis01,atsushi05} or~\cite{Hafermann14-2}.
The first steps of the following derivation were also done in~\cite{Dare94}, avoiding the introduction of a current operator.
This makes the resulting Ward identities independent of the particular form of the dispersion $\varepsilon_{\kv}$.
More importantly, this allows us to derive analogous Ward identities for quantum impurity models in Appendix~\ref{app:Wardlocal}.

The Ward identities can be viewed as sum rules for the four-point correlation function
\begin{align}
    &G^{(2),\alpha}_{\kv\kv'\qv}(\tau_1,\tau_2,\tau_3,\tau_4)=-\frac{1}{2}\sum_{\sigma_1\sigma_1'\sigma_2\sigma_2'}s^\alpha_{\sigma_1'\sigma_1}s^\alpha_{\sigma_2'\sigma_2}\label{app:gsusc_time}\\
                               &\left\langle{T_\tau c^{}_{\kv\sigma_1}(\tau_1)c^{\dagger}_{\kv+\qv,\sigma_1'}(\tau_2)c^{}_{\kv'+\qv,\sigma_2}(\tau_3)c^{\dagger}_{\kv'\sigma_2'}(\tau_4)}\right\rangle\notag,
\end{align}
which relate it to single-particle quantities.
To obtain them we examine the time derivative of this correlation function at equal times $\tau_{3}=\tau_{4}=\tau$, which allows us to express the result in terms of the time derivative of the density operators,
$\rho^\alpha_{\qv}=\sum_{\kv\sigma\sigma'}c^{\dagger}_{\kv\sigma}s^\alpha_{\sigma\sigma'} c^{}_{\kv+\qv,\sigma'}$,
\begin{align}
    &\partial_\tau\sum_{\kv'}G^{(2),\alpha}_{\kv\kv'\qv}(\tau_1,\tau_2,\tau,\tau)\label{app:gsusc_eom}\notag\\
=&\partial_\tau\frac{1}{2}\sum_{\sigma\sigma'}s^\alpha_{\sigma'\sigma}\langle T_\tau c^{}_{\kv\sigma}(\tau_1)c^{\dagger}_{\kv+\qv,\sigma'}(\tau_2)\rho^\alpha_{\qv}(\tau)\rangle\notag\\
=&\frac{1}{2}\sum_{\sigma\sigma'}s^\alpha_{\sigma'\sigma}\left\{\langle T_\tau c^{}_{\kv\sigma}(\tau_1)[\rho^\alpha_{\qv}(\tau),c^{\dagger}_{\kv+\qv,\sigma'}(\tau)]\rangle\delta_{\tau,\tau_2}\right.\notag\\
+&\langle T_\tau c^{\dagger}_{\kv+\qv,\sigma'}(\tau_2)[c^{}_{\kv\sigma}(\tau),\rho^\alpha_{\qv}(\tau)]\rangle\delta_{\tau,\tau_1}\notag\\
-&\left.\langle T_\tau c^{}_{\kv\sigma}(\tau_1)c^{\dagger}_{\kv+\qv,\sigma'}(\tau_2)[\rho^\alpha_{\qv}(\tau),H]\rangle\right\}.
\end{align}
The $\delta$-functions arise because the time-derivative does not commute with the time-ordering operator $T_\tau$.
In the last line we have replaced the time derivative of the density operators using the continuity equation $\partial_\tau\rho^\alpha=-[\rho^\alpha,H]$.
The commutators give $[\rho^\alpha_{\qv},c^{\dagger}_{\kv+\qv,\sigma}]=\sum_{\sigma'}s^\alpha_{\sigma'\sigma}c^{\dagger}_{\kv,\sigma'}$
and $[c_{\kv\sigma},\rho^\alpha_{\qv}]=\sum_{\sigma'}s^\alpha_{\sigma\sigma'}c^{}_{\kv+\qv,\sigma'}$.
We identify Green's function $G_{\kv\sigma}(\tau-\tau')\delta_{\sigma\sigma'}=-\langle T_\tau c^{}_{\kv\sigma}(\tau) c^{\dagger}_{\kv\sigma'}(\tau') \rangle$
and bring the last term in Eq.~\eqref{app:gsusc_eom} to the left-hand side (LHS) to obtain the intermediate result
\begin{align}
    &\partial_\tau\sum_{\kv'}G^{(2),\alpha}_{\kv\kv'\qv}(\tau_1,\tau_2,\tau,\tau)\label{app:ward_intermediate}\\
+&\frac{1}{2}\sum_{\sigma\sigma'}s^\alpha_{\sigma'\sigma}\langle T_\tau c^{}_{\kv\sigma}(\tau_1)c^{\dagger}_{\kv+\qv,\sigma'}(\tau_2)[\rho^\alpha_{\qv}(\tau),H]\rangle\notag\\
=&\frac{1}{2}\sum_{\sigma\sigma'}s^\alpha_{\sigma\sigma'}s^\alpha_{\sigma'\sigma}[G_{\kv+\qv,\sigma'}(\tau-\tau_2)\delta_{\tau,\tau_1}-G_{\kv\sigma}(\tau_1-\tau)\delta_{\tau,\tau_2}].\notag
\end{align}
Assuming paramagnetism, $G_\uparrow=G_\downarrow\equiv G$, we can use $\sum_{\sigma\sigma'}s^\alpha_{\sigma\sigma'}s^\alpha_{\sigma'\sigma}=2$.
We further use $[\rho^\alpha_{\qv},H_0]=\sum_{\kv'\sigma\sigma'}(\varepsilon_{\kv'+\qv}-\varepsilon_{\kv'})c^{\dagger}_{\kv'\sigma}s^\alpha_{\sigma\sigma'}c^{}_{\kv'+\qv,\sigma'}$
to separate the non-interacting from the interacting Hamiltonian,
\begin{align}
&\sum_{\kv'}(\partial_\tau+\varepsilon_{\kv'+\qv}-\varepsilon_{\kv'})G^{(2),\alpha}_{\kv\kv'\qv}(\tau_1,\tau_2,\tau,\tau)\\
+&\frac{1}{2}\sum_{\sigma\sigma'}s^\alpha_{\sigma'\sigma}\langle T_\tau c^{}_{\kv\sigma}(\tau_1)c^{\dagger}_{\kv+\qv,\sigma'}(\tau_2)[\rho^\alpha_{\qv}(\tau),H_{\text{int}}]\rangle\notag\\
=&G_{\kv+\qv}(\tau-\tau_2)\delta_{\tau,\tau_1}-G_{\kv}(\tau_1-\tau)\delta_{\tau,\tau_2}.\notag
\end{align}
In the last step we make use of the Fourier transform,
\begin{align}
    &G^{(2)}(\tau_1,\tau_2,\tau_3,\tau_4)\\
    =&\sum_{\nu\nu'\omega}G^{(2)}_{\nu\nu'\omega}e^{-\imath[\nu\tau_1-(\nu+\omega)\tau_2+(\nu'+\omega)\tau_3-\nu'\tau_4]},\notag
\end{align}
introduce the short notation $k=(\kv,\nu)$, $q=(\qv,\omega)$, and substitute the generalized susceptibility $X^{\alpha}_{kk'q}=G^{(2),\alpha}_{kk'q}+2\beta G_kG_{k'}\delta_q\delta_\alpha$ to obtain the Ward identities
\footnote{The term $\propto\delta_{q=(\qv,\omega),(0,0)}$ does not contribute to the Ward identities.},
\begin{align}
G_{k+q}-G_k=&\sum_{k'}X^{\alpha}_{kk'q}\left[\varepsilon_{k'+q}-\varepsilon_{k'}-\imath\omega\right]\notag\\
            &+\frac{1}{2}\sum_{\sigma\sigma'}s^\alpha_{\sigma'\sigma}\langle c^{}_{k\sigma}c^{\dagger}_{k+q,\sigma'}[\rho^\alpha_{q},H_{\text{int}}]\rangle.\label{app:ward_general_lattice}
\end{align}
The first term on the right-hand-side (RHS) can be recognized as the contribution from the non-interacting Hamiltonian, whereas the second contribution originates from the interaction.

In the Hubbard model the interaction conserves the densities $\rho^\alpha$, that is $[\rho^\alpha,U n_{\uparrow}n_{\downarrow}]=0$ for $\alpha=0,x,y,z$. Hence the contribution of the interaction to the current in the Ward identities, the second line of~\eqref{app:ward_general_lattice}, vanishes.
This allows to recast Eq.~\eqref{app:ward_general_lattice} into a relation between the irreducible vertex $\Gamma^{}$, Green's function $G$ and the self-energy $\Sigma$:
The generalized susceptibility $X^\alpha_{kk'q}$ is related to the irreducible vertex $\Gamma^{\alpha}_{kk'q}$ via the integral equation
$X_{kk'q}^{\alpha} =G_{k}G_{k+q}\left[\beta \delta_{kk'}-\sum_{k''}\Gamma_{kk''q}^{\alpha}X^\alpha_{k''k'q}\right]$.
We insert this relation into Eq.~\eqref{app:ward_general_lattice} and divide by $G_kG_{k+q}$.
Using $\varepsilon_{k+q}-\varepsilon_{k}-\imath\omega=[G^0_k]^{-1}-[G^0_{k+q}]^{-1}$ and $[G^0_k]^{-1}-G_k^{-1}=\Sigma_k$ one has
\begin{align}
    \Sigma_{k+q}-\Sigma_k = -\sum_{k'}\Gamma_{kk'q}^{\alpha}\sum_{k''}X^{\alpha}_{k'k''q}\left[\varepsilon_{k''+q}-\varepsilon_{k''}-\imath\omega\right].\notag
\end{align}
Since $[\rho^\alpha,H_{\text{int}}]=0$, one can in turn insert the Ward identities~\eqref{app:ward_general_lattice} on the RHS to obtain the desired relation,
\begin{align}
    \Sigma_{k+q}-\Sigma_k = -\sum_{k'}\Gamma_{kk'q}^{\alpha}\left[G_{k'+q}-G_{k'}\right]\label{app:ward}.
\end{align}

\section{Susceptibility asymptote and $f$-sum rule}
Assuming that the interaction does not contribute to the currents, i.e., letting $[\rho^\alpha,H_{\text{int}}]=0$ in the Ward identities~\eqref{app:ward_general_lattice},
we derive the $(\imath\omega)^{-2}$ coefficient of the lattice susceptibility of locally conserving approximations.
We prove that in this case local conservation implies the $f$-sum rule.

\subsection{Susceptibility asymptote}
\label{app:asymptotefromward}
We recognize that the second line of~\eqref{app:ward_general_lattice} vanishes due to $[\rho^\alpha,H_{\text{int}}]=0$. Summing over $k$ yields zero on the LHS and we are left with
\begin{align}
0=\sum_{kk'}X^{\alpha}_{kk'q}\left[\varepsilon_{k'+q}-\varepsilon_{k'}-\imath\omega\right]. \label{app:wardingredient}
\end{align}
We use $X^{\alpha}_{q}=2\sum_{kk'}X^{\alpha}_{kk'q}$ to arrive at an exact expression for the susceptibility,
\begin{align}
    \imath\omega X^{\alpha}_q
    = 2\sum_{kk'}X^{\alpha}_{kk'q}\left[\varepsilon_{k'+q}-\varepsilon_{k'}\right].
  \label{app:sumrule_anyomega}
\end{align}
This implies for the homogeneous limit, $\qv\to 0$ for finite frequency, that
\begin{align}
    X^{\alpha}_{\qv=0,\omega\neq0}=-\langle(\rho^\alpha_{\qv=0,\omega\neq0})^2\rangle=0.\label{app:static_homogenous_susc}
\end{align}
$\rho^\alpha_{\qv=0}=\sum_i\rho^\alpha_{i}$ is the operator of total charge (spin) which is conserved if the continuity equations hold globally, that is $\partial_\tau\rho^\alpha_{\qv=0,\tau}=0$.
Thus, $\omega\rho^\alpha_{\qv=0,\omega}=0$ or~\eqref{app:static_homogenous_susc} reflects that the approximation to $X^\alpha_{\qv\omega}$ conserves the total charge or spin, respectively.
Therefore, as expected, local conservation implies global conservation.

To obtain the $(\imath\omega)^{-2}$ coefficient of $X^\alpha_{q}$, we expand $X^\alpha_{kk'q}=G_{k}G_{k+q}\left[\beta \delta_{kk'}-\sum_{k''}\Gamma_{kk''q}^{\alpha}X_{k''k'q}^\alpha\right]$ on the RHS of Eq.~\eqref{app:sumrule_anyomega} to order $\mathcal{O}(\omega^{-1})$:
According to the following Eq.~\eqref{eq:app:highfrequencyexpansionX}, the interacting bubble $G_{k}G_{k+q}$ decays at least as $\omega^{-1}$.
Since we can treat $\Gamma^{}$ as a constant at $\omega\rightarrow\infty$,
vertex corrections to $X^\alpha_{kk'q}$ are negligible at order $\mathcal{O}(\omega^{-1})$ and one is left with
\begin{align}
    X^{\alpha}_{kk'q}
    =&\beta G_{k}G_{k+q}\delta_{kk'}+\mathcal{O}(\omega^{-2})\notag\\
    =&\frac{\beta}{\imath\omega}\frac{G_{k}-G_{k+q}}{1+(\varepsilon_k-\varepsilon_{k+q}+\Sigma_k-\Sigma_{k+q})/\imath\omega}\delta_{kk'}+\mathcal{O}(\omega^{-2})\notag \\
    =&\frac{\beta}{\imath\omega}(G_{k}-G_{k+q})\delta_{kk'}+\mathcal{O}(\omega^{-2}).\label{eq:app:highfrequencyexpansionX}
\end{align}
The Green's function $G_{k+q}$ in the last line contributes since the term can be of order $\mathcal{O}(1)$ for $k\approx -q$.
We return to the usual frequency and momentum notation, $k=(\kv,\nu)$, $q=(\qv,\omega)$, and insert~\eqref{eq:app:highfrequencyexpansionX} into the RHS of \eqref{app:sumrule_anyomega}. This yields the asymptotic coefficient of $X^\alpha_{\qv\omega}$ at order $\mathcal{O}(\omega^{-2})$:
\begin{align}
    &\lim\limits_{\omega\rightarrow\infty}(\imath\omega)^2 X^\alpha_{\qv\omega}\notag\\
    =&\lim\limits_{\omega\rightarrow\infty}2\sum_{\nu\kv}(G_{\kv\nu}-G_{\kv+\qv,\nu+\omega})\left[\varepsilon_{\kv+\qv}-\varepsilon_{\kv}\right]\notag\\
    =&2\sum_{\nu\kv}(G_{\kv\nu}\varepsilon_{\kv+\qv}+G_{\kv+\qv,\nu}\varepsilon_{\kv}-2G_{\kv\nu}\varepsilon_{\kv})\notag\\
    =&\sum_{\kv\sigma}\av{n_{\kv\sigma}}(\varepsilon_{\kv+\qv}+\varepsilon_{\kv-\qv}-2\varepsilon_{\kv}).
  \label{app:sumrule_largeomega}
\end{align}
Here we have used $\sum_\nu G_{\kv\nu\sigma}e^{\imath\nu0^+}=\av{n_{\kv\sigma}}$
\footnote{The factor $e^{\imath\nu0^+}$ was left out for readability. It needs to be inserted in~\eqref{app:ward_general_lattice} before summation over $k$.}.
We show below that this equation is in fact the $f$-sum rule.
For the local susceptibility $X^\alpha_{\text{loc},\omega}=\sum_{\qv}X^\alpha_{\qv\omega}$ it follows using $\sum_{\kv}\varepsilon_{\kv}=0$
and $E_\text{kin}=\sum_{\kv\sigma}\av{n_{\kv\sigma}}\varepsilon_{\kv}$, that
\begin{equation}
\lim\limits_{\omega\rightarrow\infty}(\imath\omega)^2X^\alpha_{\text{loc},\omega}=-2E_\text{kin},
\end{equation}
which is used several times in the main text.

Lastly, we simplify the asymptotic coefficient on the RHS of Eq.~\eqref{app:sumrule_largeomega} for a square lattice with nearest-neighbor hopping, $\varepsilon_\kv=-2t[\cos k_x+\cos k_y]$.
In this case one can use goniometric equalities to extract the dependence on $\qv$ from Eq.~\eqref{app:sumrule_largeomega} as
\begin{align}
    & \lim\limits_{\omega\rightarrow\infty}(\imath\omega)^2 X^\alpha_{\qv\omega}= 2\sum_{\kv\sigma}\langle n_{\kv\sigma}\rangle(\varepsilon_{\kv+\qv}-\varepsilon_{\kv}) \notag \\
 =&
 -4t\sum_{\kv\sigma}\langle n_{\kv\sigma}\rangle\sum_{i=x,y}\left(\cos k_i(\cos q_i-1)-\sin k_i\sin q_i\right) \notag \\
 =&
 -4t
 \sum_{i=x,y}
 \left(\cos q_i-1\right)
 \sum_{\kv\sigma}
 \langle n_{\kv\sigma}\rangle 
 \cos k_i
 \notag \\
 =&
 \sum_{i=x,y}
 \left(\cos q_i-1\right)
 \sum_{\kv\sigma}
 \langle n_{\kv\sigma}\rangle 
 \varepsilon_{\kv}
 \notag \\ 
 =& \left( \cos q_x+\cos q_y - 2 \right)\,  E_\text{kin}.\label{eq:app:asymptote:kinetic}
\end{align}
It was used in the first line that $\varepsilon_{-\kv}=\varepsilon_{\kv}$, in the second line that $\sum_{\kv} \av{n_{\kv}} \sin k_i =0$  and in the third line that 
$\sum_{\kv}\langle n_{\kv}\rangle \cos k_x=\sum_{\kv}\langle n_{\kv}\rangle \cos k_y$, all valid by symmetry of the lattice.

Fig.~\ref{fig:fsumrule_test} numerically illustrates the validity of Eqs.~\eqref{app:sumrule_largeomega} and~\eqref{eq:app:asymptote:kinetic} by plotting both sides of the equation (for the LHS, we take large but finite values of $\omega$) in DMFT and two-particle self-consistent DMFT (cf. section~\ref{section:scdb}).
At the M-point, the RHS equals $-4E_\text{kin}$, which is marked by dashed horizontal lines.
This value is larger in DMFT and given by $-8\sum_{\nu}\Delta_{\nu}g_{\nu}$, since the kinetic energy can be computed from the impurity (cf. sections~\ref{sec:DMFT} and~\ref{section:sz-coupling}).

\subsection{$f$-sum rule}
\label{app:fsumrule_derivation}
\begin{figure}
 \includegraphics[]{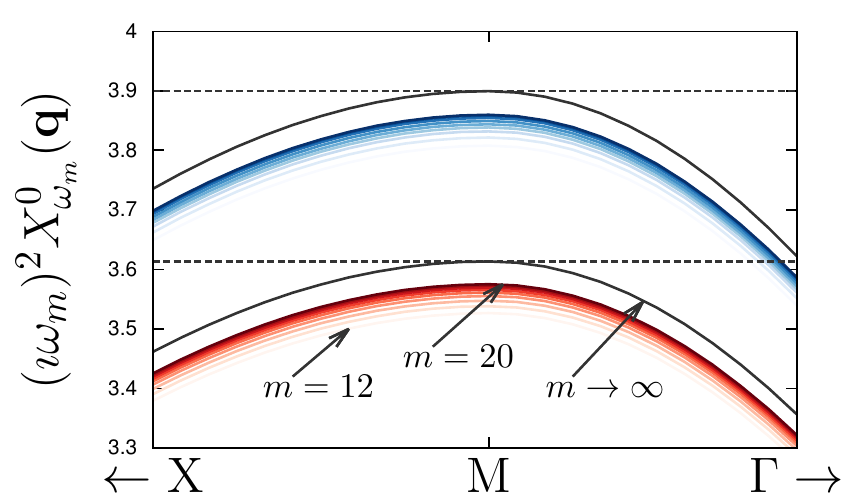}
 \caption{\label{fig:fsumrule_test} (Color online)
     Convergence of $(\imath\omega_m)^2X^0_{\qv\omega_m}$ to the RHS of the $f$-sum rule~\eqref{app:fsumrule_sine}.
     The figure shows the charge susceptibility in DMFT (blue) and two-particle self-consistent DMFT (red).
     Darker colors indicate a larger Matsubara index $m\leq20$.
     Solid black curves mark the analytical expression for $m\rightarrow\infty$, Eq.~\eqref{eq:app:asymptote:kinetic}.
     Dashed lines indicate the value $-4E_\text{kin}$, (half-filling, $U=6, \beta=2$).}
\end{figure}
The so-called $f$-sum rule (see, for example,~\cite{Vilk97}),
\begin{align}
    -&4\lim\limits_{\eta \rightarrow 0}\frac{1}{\beta}\sum_{n>0}^\infty\omega_n\sin(\eta\omega_n)X^\alpha_{\qv\omega_n}\notag\\
    &=\sum_{\kv\sigma}\av{n_{\kv\sigma}}(\varepsilon_{\kv+\qv}+\varepsilon_{\kv-\qv}-2\varepsilon_{\kv}),\label{app:fsumrule_sine}
\end{align}
is a relation between the 2P response (LHS) and 1P quantities (RHS).
To avoid confusion, we write the factor $\beta^{-1}$ in front of the sum explicitly.

The Ward identities of the Hubbard model imply the $f$-sum rule~\cite{Vilk97}.
It was also mentioned in~\cite{Vilk97} that the LHS of Eq.~\eqref{app:fsumrule_sine} is entirely determined by the leading $(\imath\omega)^{-2}$ coefficient of $X^\alpha_{\qv\omega}$.
We can see this directly by comparing~\eqref{app:sumrule_largeomega} with the RHS of Eq.~\eqref{app:fsumrule_sine}.
Here we show that the sum on the LHS of~\eqref{app:fsumrule_sine} indeed singles out the leading coefficient in the high-frequency expansion of the susceptibility.

The limit $\eta\rightarrow 0$ of every summand on the LHS of Eq.~\eqref{app:fsumrule_sine} is zero whereas the limit of the sum is not.
The LHS is convergent but not necessarily absolutely convergent. We drop the labels $\alpha$ and $\qv$ temporarily and expand $X^\alpha_{\qv\omega}$ for large frequencies.
Since $X(\omega_n)=X(-\omega_n)$, only even powers of $\omega_n$ appear in the expansion, $X(\omega_n) = \sum_{k=1}^\infty a_{2k} \omega_n^{-2k}$.
Inserting this expression into the $f$-sum rule~\eqref{app:fsumrule_sine}, one has on the LHS
\begin{align}
    &-4\sum_{k=1}^\infty a_{2k} \lim_{\eta\rightarrow 0}\frac{1}{\beta}\sum_{n=1}^\infty  \frac{\sin(\eta \omega_n)}{ \omega_n^{2k-1}}.
\end{align}
We hence need to evaluate
\begin{align}
    \lim_{\eta\rightarrow 0}\sum_{n=1}^\infty \frac{\sin(\eta n)}{n^{2k-1}}&=
  \begin{cases} 
      \lim_{\eta\rightarrow 0}\frac{\pi-\eta}{2}=\frac{\pi}{2} & \text{if } k=1,\\
      0 & \text{if } k>1.
  \end{cases}\label{eq:fsum:conditionalconvergence}
\end{align}
For $k=1$, the limit $\eta\rightarrow 0$ and the summation over $n$ must not be interchanged, since $\sum^\infty_{n=1}\left|\sin(\eta n) \right|/n$ diverges for $0<\eta<\pi$.
For the higher order coefficients, the sum in Eq.~\eqref{eq:fsum:conditionalconvergence} is absolutely convergent, the limit and the sum can be interchanged, leading to zero.
Correspondingly, the LHS of~\eqref{app:fsumrule_sine} becomes
\begin{align}
&-4a_2\lim_{\eta\rightarrow 0}\frac{1}{\beta}\sum_{n=1}^\infty  \frac{\sin(\eta n)}{ 2\pi n/\beta} = -a_2.
\end{align}

As expected, only the $(i\omega)^{-2}$ coefficient $-a_2=\lim\limits_{\omega\rightarrow\infty}(\imath\omega)^2 X(\imath\omega)$ determines the LHS of the $f$-sum rule.
As we have seen in the last paragraph, the $f$-sum rule follows from local conservation.

The above result is useful because a straightforward numerical evaluation of the $f$-sum rule~\eqref{app:fsumrule_sine} suffers from oscillatory behavior of the LHS with the cutoff frequency.
On the other hand, $(\imath\omega)^2 X^\alpha_{\qv\omega}$ approaches the limit $\omega\rightarrow\infty$ smoothly, which is illustrated in Fig.~\ref{fig:fsumrule_test}.
For an accurate extrapolation to this limit, one needs to account for the effect of finite frequency cutoffs in the vertex corrections to $X^\alpha_{\qv\omega}$.

\section{Asymptote of the impurity susceptibility}
We determine the $(\imath\omega)^{-2}$ asymptote of the impurity susceptibility $\chi^\alpha$.
It can be shown from the Lehmann representation of $\chi^\alpha$ that under paramagnetism its asymptote takes on the following form:
\begin{align}
    \lim\limits_{\omega\rightarrow\infty}(\imath\omega)^2\chi^{\alpha}_\omega = \left\langle\left[[\bar{\rho}^\alpha,H_{\text{imp}}],\bar{\rho}^\alpha\right]\right\rangle.\label{app:lehmann_asymptote}
\end{align}

\subsection{Impurity Hamiltonian}
\label{app:impurity:hamiltonian}

In order to evaluate~\eqref{app:lehmann_asymptote} in the Bose-Fermi-Kondo model, Eq.~\eqref{BFK}, we need to use its Hamiltonian formulation,
\begin{align}
  H_{\text{imp}}=H_{\text{at}}+H^0_\Delta+H_{\Delta}+H^{0}_\Lambda+H_\Lambda.\label{bose_anderson}
\end{align}
The first three components,
\begin{align}
  H_{\text{at}}=&-\tilde{\mu} n+\tilde{U} n_{\uparrow}n_{\downarrow},\notag\\
  H^0_{\Delta}=&\sum_{\vect{k}\sigma}\epsilon_{\vect{k}}f^\dagger_{\vect{k}\sigma}f^{}_{\vect{k}\sigma},\notag\\
  H_{\Delta}=&\sum_{\vect{k}\sigma}(v^{}_{\vect{k}}c^\dagger_{\sigma}f^{}_{\vect{k}\sigma}+v^*_{\vect{k}}f^\dagger_{\vect{k}\sigma}c^{}_{\sigma}),\label{fermion_bath}
\end{align}
are the constituents of the Anderson impurity model, where a correlated impurity $H_{\text{at}}$ is coupled to a non-interacting bath $H^0_{\Delta}$
via the hybridization $H_{\Delta}$. In the BFK we further have the bosonic contributions
\begin{align}
  H^0_{\Lambda}=&\sum_\alpha H^{0,\alpha}_\Lambda,\;H_{\Lambda}=\sum_\alpha H^{\alpha}_\Lambda,\notag\\
  H^{0,\alpha}_{\Lambda}=&\sum_{\vect{q}}\Omega^\alpha_{\vect{q}}(b^\alpha_{\vect{q}})^\dagger b^{\alpha}_{\vect{q}},\notag\\
  H^\alpha_{\Lambda}=&\sum_{\vect{q}}w^\alpha_{\vect{q}}\bar{\rho}^\alpha\phi^\alpha_{\vect{q}},\;\phi^\alpha_{\vect{q}}=((b^\alpha_{\vect{q}})^\dagger+b^{\alpha}_{\vect{q}}),\notag
\end{align}
these couple the correlated site to bosonic baths $H^0_{\Lambda}$ via the density-boson interactions $H_{\Lambda}$.
Labeling the quantum numbers of the fermionic and bosonic baths with $\vect{k}$ and $\vect{q}$, respectively, the situation described by the BFK is summarized as follows:
The matrix elements $v^{}_{\vect{k}}$ ($w^\alpha_{\vect{q}}$) couple the correlated fermions $c^\dagger,c$ (density $\rho^\alpha$) to a bath of non-interacting fermions $f^\dagger,f$ (bosons $\phi^\alpha$) with the spectrum $\epsilon_{\vect{k}}$ ($\Omega^\alpha_{\vect{q}}$).

Integrating out the baths in the path integral formalism yields the effective action of the BFK (\ref{eq:AIM}), with the hybridization functions
\begin{align}
  \Delta_{\nu\sigma}=&\sum_{\vect{k}}|v_{\vect{k}}|^2\mathcal{G}_{\vect{k}\sigma\nu}\label{discretedelta},\\
  \tilde{\Lambda}^\alpha_\omega=&\sum_{\vect{q}}(w^\alpha_{\vect{q}})^2\mathcal{D}^\alpha_{\vect{q}\omega}.\label{discretelambda}
\end{align}
Here $\mathcal{G}_{\vect{k}\nu}=1/(\imath\nu-\varepsilon_{\vect{k}})$ and $\mathcal{D}^\alpha_{\vect{q}\omega}=-2\Omega^\alpha_{\vect{q}}/(\omega^2+(\Omega^\alpha_{\vect{q}})^2)$
denote the bath Green's functions.
The self-consistent interaction $\Lambda^\alpha=\Lambda^\alpha_\infty+\tilde{\Lambda}^\alpha_\omega$ has a constant part $\Lambda^\alpha_\infty$ and a dynamic part $\tilde{\Lambda}^\alpha_\omega$,
which has to vanish at $\omega\rightarrow\infty$ (cf.~\eqref{discretelambda}).
The constant parts are absorbed into the Hubbard interaction $\tilde{U}=U+\Lambda^0_\infty-f\Lambda^z_\infty$ where $f=1$ in the Ising-type coupling
and $f=3$ in the Heisenberg-type coupling, cf. Sec.~\ref{sec:2psc}.
The interactions $\Lambda^\alpha$ change the chemical potential $\tilde{\mu}=\mu+\mu_{\text{shift}}$, which is shifted such that half filling is obtained at $\tilde{\mu}=\tilde{U}/2$.

\subsection{Susceptibility asymptotes}
\label{app:asymptotics:impurity}

To determine the asymptotic coefficient, Eq.~\eqref{app:lehmann_asymptote}, we need to calculate
\begin{align*}
  C^\alpha=&\left\langle\left[[\bar{\rho}^\alpha,H_{\text{imp}}],\bar{\rho}^\alpha\right]\right\rangle.
\end{align*}
Only the operators $H_\Delta, H_\Lambda$ contribute to this expression:
\begin{align}
  C^\alpha=& C_{\Delta} + \tilde{C}^\alpha_{\Lambda}\label{total_imp_asymptote},\\
  C_{\Delta}=&\left\langle\left[[\bar{\rho}^\alpha,H_\Delta],\bar{\rho}^\alpha\right]\right\rangle=-\langle H_\Delta\rangle,\\
  \tilde{C}^\alpha_{\Lambda}=&\sum_\gamma\left\langle\left[[\bar{\rho}^\alpha,H^\gamma_\Lambda],\bar{\rho}^\alpha\right]\right\rangle=
  \begin{cases} 
     0 & \text{if } \alpha=0,\\
    \sum_{\gamma\neq\alpha} C^\gamma_{\Lambda}& \text{if } \alpha=x,y,z,
  \end{cases}\nonumber\\
  C^\gamma_{\Lambda}=& -4\langle H^\gamma_{\Lambda}\rangle.\label{bosonic_imp_asymptote_cases}
\end{align}
We calculate these expectation values from the impurity action. To this end we denote the action of the Hamiltonian $H^0_{\text{imp}}=H_{\text{at}}+H^0_{\Delta}+H^0_{\Lambda}$ as $S^0$
and add sources $J, J^\alpha$ to the actions corresponding to the operators $H_{\Delta}$ and $H_{\Lambda}=\sum_\alpha H^\alpha_\Lambda$, respectively:
\begin{align*}
  S^{\Delta}(J)=&\int_0^\beta d\tau\left\{\sum_{\vect{k}\sigma}[v^{}_{\vect{k}}+J_{\vect{k}\sigma\tau}]c^*_{\sigma\tau}f^{}_{\vect{k}\sigma\tau}+v^*_{\vect{k}}f^*_{\vect{k}\sigma\tau}c^{}_{\sigma\tau}\right\},\\
  S^{\Lambda}(J^\alpha)=&\int_0^\beta d\tau\sum_{\vect{q}\alpha}[w^\alpha_{\vect{q}}+J^\alpha_{\vect{q}\tau}]\bar{\rho}^\alpha_\tau\phi^\alpha_{\vect{q}\tau}.
\end{align*}
The expectation values can then be obtained as functional derivatives ($S^{\Delta,\Lambda}_\text{imp}=S^0+S^{\Delta}+S^{\Lambda}$),
\begin{align}
    &-\langle H_{\Delta}\rangle = -\sum_{\vect{k}\sigma}\left\{v^{}_{\vect{k}}\langle c^\dagger_{\sigma}f^{}_{\vect{k}\sigma}\rangle+v^*_{\vect{k}}\langle f^\dagger_{\vect{k}\sigma}c^{}_{\sigma}\rangle\right\},\label{app:exphdelta}\\
  &\langle c^\dagger_{\sigma}f^{}_{\vect{k}\sigma}\rangle = -\left.\frac{1}{\mathcal{Z}}\frac{\delta}{\delta J^{}_{\vect{k}\sigma\tau=0}}\right|_{J=0}
  \int\mathcal{D}[c,f,b]e^{-S^{\Delta,\Lambda}_\text{imp}},\notag
\end{align}
where $\mathcal{Z}$ is the grand partition sum, and
\begin{align}
    &-\langle H^\gamma_{\Lambda}\rangle=-\sum_{\vect{q}}w^\gamma_{\vect{q}}\left\langle\bar{\rho}^\gamma\phi^\gamma_{\vect{q}}\right\rangle\label{app:exphgamma},\\
  &\left\langle\bar{\rho}^\gamma\phi^\gamma_{\vect{q}}\right\rangle = -\left.\frac{1}{\mathcal{Z}}\frac{\delta}{\delta J^{\gamma}_{\vect{q}\tau=0}}\right|_{J^\gamma=0}
  \int\mathcal{D}[c,f,b]e^{-S^{\Delta,\Lambda}_\text{imp}}.\notag
\end{align}
The next step is to integrate out the fermionic and bosonic baths, $\int\mathcal{D}[c,f,b]e^{-S^{\Delta,\Lambda}_\text{imp}}=\mathcal{Z}_{f,b}\int\mathcal{D}[c]e^{-S'_\text{imp}}$.
This gives rise to retarded couplings of $c_\tau^*,c_{\tau'}$ and $\rho^\alpha_\tau,\rho^\alpha_{\tau'}$ via the respective bath Green's functions $\mathcal{G}_{\tau-\tau'}$ and $\mathcal{D}_{\tau-\tau'}$ (see below~\eqref{discretelambda},\footnote{For readability, expectation values are still marked with the same brackets after integrating out the baths.}),
\begin{align*}
  &S'_\text{imp}(J,J^\alpha)=\int_0^\beta d\tau \left\{c^*_{\sigma\tau}(\partial_\tau-\tilde{\mu})c_{\sigma\tau}+Un_{\uparrow\tau}n_{\downarrow\tau}\right\}\\
  &+\int_0^\beta d\tau d\tau' \sum_{\vect{k}\sigma}[v^{}_{\vect{k}}+J_{\vect{k}\sigma\tau}]c^*_{\sigma\tau}\mathcal{G}_{\vect{k}\sigma,\tau-\tau'}c^{}_{\sigma\tau'}v^*_{\vect{k}}\\
  &+\frac{1}{2}\int_0^\beta d\tau d\tau' \sum_{\vect{q}\alpha}[w^\alpha_{\vect{q}}+J^\alpha_{\vect{q}\tau}]\bar{\rho}^\alpha_\tau\mathcal{D}_{\vect{q},\tau-\tau'}^{\alpha}\bar{\rho}^\alpha_{\tau'}[w^\alpha_{\vect{q}}+J^\alpha_{\vect{q}\tau'}].
\end{align*}
Performing the derivatives with respect to the sources $J, J^\alpha$ relates the desired expectation values $\langle c^\dagger_{\sigma}f^{}_{\vect{k}\sigma}\rangle$ and $\left\langle\bar{\rho}^\gamma\phi^\gamma_{\vect{q}}\right\rangle$ to the correlation functions $g$ and $\chi$,
\begin{align}
  \langle c^\dagger_{\sigma}f^{}_{\vect{k}\sigma}\rangle =&v^*_{\vect{k}}\int_0^\beta d\tau' \mathcal{G}_{\vect{k}\sigma,0-\tau'}\langle c^*_{\sigma\tau=0}c^{}_{\sigma\tau'}\rangle\notag\\
  =&v^*_{\vect{k}}\sum_\nu\mathcal{G}_{\vect{k}\sigma\nu}g_{\nu\sigma}\label{app:av_rf},\\
  \left\langle\bar{\rho}^\gamma\phi^\gamma_{\vect{q}}\right\rangle =& w^\gamma_{\vect{q}}\int_0^\beta d\tau' \mathcal{D}_{\vect{q},0-\tau'}^{\gamma}\langle \bar{\rho}^\gamma_{\tau=0}\bar{\rho}^\gamma_{\tau'}\rangle\notag\\
  =&-w^\gamma_{\vect{q}}\sum_\omega\mathcal{D}_{\vect{q},\omega}^{\gamma}\chi^\gamma_\omega\label{app:av_rb}.
\end{align}
Here we have identified $-\langle c^{}_{\sigma\tau'}c^*_{\sigma\tau=0}\rangle=g_{\sigma\tau'}$ and $-\langle \bar{\rho}^\gamma_{\tau'}\bar{\rho}^\gamma_{\tau=0}\rangle=\chi^\gamma_{\tau'}$.
Similarly,
\begin{align*}
  \langle f^\dagger_{\sigma}c^{}_{\vect{k}\sigma}\rangle =&v_{\vect{k}}\int_0^\beta d\tau \mathcal{G}_{\vect{k}\sigma,\tau-0}\langle c^*_{\sigma\tau}c^{}_{\sigma\tau'=0}\rangle\\
  =&v_{\vect{k}}\sum_\nu\mathcal{G}_{\vect{k}\sigma\nu}g_{\nu\sigma}.
\end{align*}
Inserting the expectation values into Eqs.~\eqref{app:exphdelta} and~\eqref{app:exphgamma}, we finally determine the asymptotic coefficients,
\begin{align*}
  C_{\Delta}=&-\langle H_\Delta\rangle = -2\sum_{\vect{k}\sigma\nu}|v_{\vect{k}}|^2\mathcal{G}_{\vect{k}\sigma\nu}g_{\nu\sigma},\\
  C^\gamma_{\Lambda}=&-4\langle H^\gamma_\Lambda\rangle=4\sum_{\vect{q}\omega}(w^\gamma_{\vect{q}})^2\mathcal{D}_{\vect{q},\omega}^{\gamma}\chi^\gamma_\omega.
\end{align*}
Using the definitions of the bath Green's functions, Eqs.~\eqref{discretedelta} and~\eqref{discretelambda}, we conclude
\begin{align}
     C_{\Delta}=&-2\sum_{\nu\sigma}\Delta_{\nu\sigma}g_{\nu\sigma}\label{imp_cf},\\
     C_{\Lambda}^\gamma=&4\sum_{\omega}\tilde{\Lambda}^\gamma_\omega \chi^\gamma_\omega.\label{imp_cb}
\end{align}
\begin{figure}
    \centering
  \begin{minipage}[b]{0.235\textwidth}
      \includegraphics[width=\textwidth]{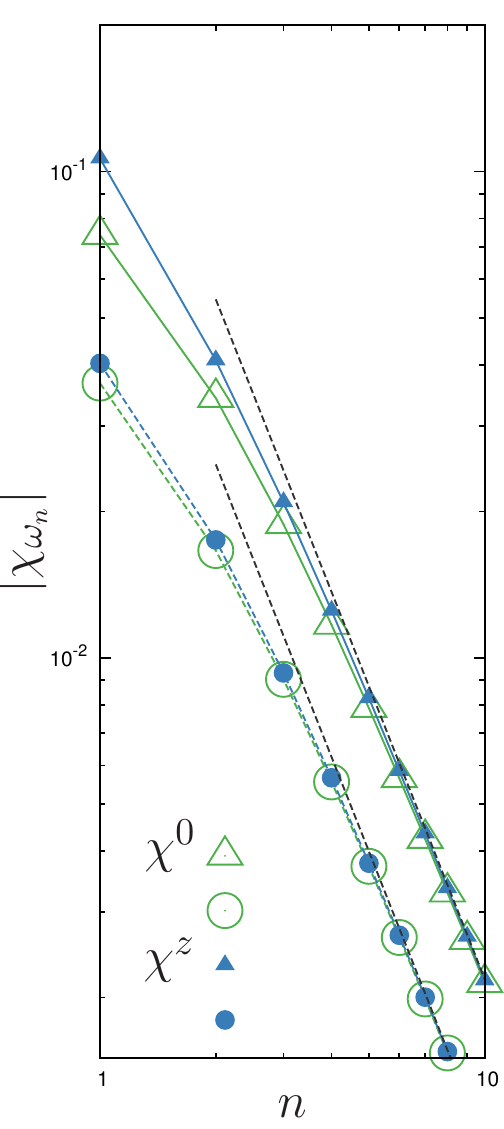}
  \end{minipage}
\hfill
  \begin{minipage}[b]{0.235\textwidth}
      \includegraphics[width=\textwidth]{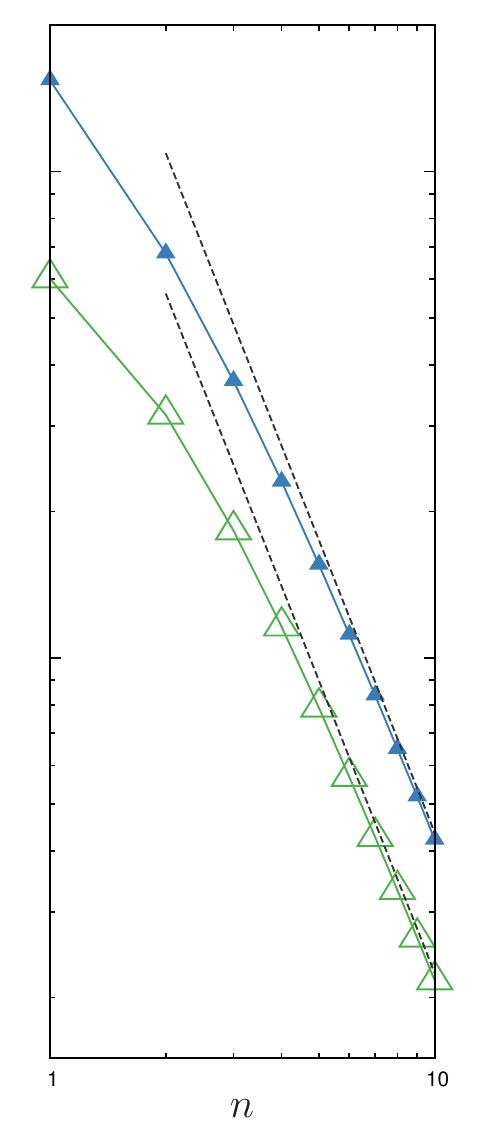}
  \end{minipage}

  \caption{(Color online) Impurity susceptibility $\chi^{\alpha}$ of the Bose-Fermi-Kondo model (symbols) and its exact high-frequency asymptote (dashed lines, see text).
   Left panel: Ising-type coupling, $\chi^0$ (green) and $\chi^z$ (blue) approach the same asymptote, (triangles: $\langle n\rangle=0.54$, circles: $\langle n\rangle=0.17$).
   Right panel: Heisenberg-type coupling, $\chi^0$ and $\chi^z$ approach different asymptotes. (Colors as in the left panel. $\beta=2$, $U=6$ in both panels.)}\label{fig:doped:chi:imp:asymptote}
\end{figure}

We examine this result for the Bose-Fermi-Kondo model~\eqref{bose_anderson} in two cases: (i) Ising-type coupling: Finite retarded interactions
$\tilde{\Lambda}^0, \tilde{\Lambda}^z$ in the density-type channels and $\tilde{\Lambda}^x=\tilde{\Lambda}^y=0$ in the transversal spin channels.
Collecting Eqs.~\eqref{total_imp_asymptote},~\eqref{bosonic_imp_asymptote_cases},~\eqref{imp_cf}, and~\eqref{imp_cb},
we find that the susceptibilities of the BFK assume the following asymptotic behavior for large $\omega$:
\begin{align}
  \lim\limits_{\omega\rightarrow\infty}(\imath\omega)^2\chi^{0,z}_\omega =&-2\sum_{\nu\sigma}\Delta_{\nu\sigma}g_{\nu\sigma},\label{app:zasymptote}\\
  \lim\limits_{\omega\rightarrow\infty}(\imath\omega)^2\chi^{x,y}_\omega =&-2\sum_{\nu\sigma}\Delta_{\nu\sigma}g_{\nu\sigma}\label{app:xyasymptote}
    +4\sum_{\omega'}\tilde{\Lambda}^z_{\omega'} \chi^z_{\omega'}.
\end{align}
The Ising-type coupling is used in the two-particle self-consistent DMFT in section~\ref{section:scdb}.
Hence in this approximation $\chi^0$ and $\chi^z$ approach the asymptote given in Eq.~\eqref{app:zasymptote}, which is demonstrated in the left panel of Fig.~\ref{fig:doped:chi:imp:asymptote}.
Results for two different fillings $\langle n\rangle$ are shown.

(ii) Heisenberg-type coupling: Finite retarded interactions in all channels, $\tilde{\Lambda}^0,\tilde{\Lambda}^x=\tilde{\Lambda}^y=\tilde{\Lambda}^z=\tilde{\Lambda}'$:
\begin{align}
  \lim\limits_{\omega\rightarrow\infty}(\imath\omega)^2\chi^0_\omega =&-2\sum_{\nu\sigma}\Delta_{\nu\sigma}g_{\nu\sigma},\label{app:chasymptote}\\
  \lim\limits_{\omega\rightarrow\infty}(\imath\omega)^2\chi^z_\omega =&-2\sum_{\nu\sigma}\Delta_{\nu\sigma}g_{\nu\sigma}
    +4\hspace*{-0.3cm}\sum_{\omega',\alpha=x,y}\hspace*{-0.3cm}\tilde{\Lambda}^\alpha_{\omega'} \chi^\alpha_{\omega'}.\label{app:xyzasymptote}
\end{align}
The asymptotes of $\chi^x$ and $\chi^y$ may be obtained by permuting the labels $x,y,z$ in Eq.~\eqref{app:xyzasymptote}.

We compare Eqs.~\eqref{app:chasymptote} and~\eqref{app:xyzasymptote} with numerical results in the right panel of Fig.~\ref{fig:doped:chi:imp:asymptote}.
We used the CTQMC solver presented in~\cite{Otsuki13} to calculate the impurity susceptibility $\chi^\alpha$ in the Bose-Fermi-Kondo model in Heisenberg-type coupling, $\tilde{\Lambda}'\neq 0$, with a conducting bath $\Delta$ at half-filling. 

\section{Migdal formula for the double occupancy}
\label{app:migdal}
In the Hubbard model, the Migdal formula $d=\av{n_\up n_\dn} = \text{Tr}(G\Sigma)/2U$ can be used to calculate the double occupancy from the potential energy $Ud$~\cite{Galitskii58}.
In the Bose-Fermi-Kondo model~\eqref{bose_anderson}, the potential energy is modified due to retarded interactions $\Lambda_\omega=\Lambda_\infty+\tilde{\Lambda}_\omega$.
Their constant parts cause a shift $U\rightarrow\tilde{U}=U+\Lambda^0_\infty-f\Lambda^z_\infty$. In the following, we determine the effect of the dynamic part $\tilde{\Lambda}_\omega$.
We start from the equation of motion (EOM) of the impurity Green's function $g_{-\tau}=-\av{T_\tau c_{-\tau\sigma}c^\dagger_\sigma}$,
\begin{align}
    &-\partial_\tau g_{-\tau\sigma} = -\delta(\tau)\av{(c_\sigma,c^\dagger_\sigma)}+\av{T_\tau\left[c_{-\tau,\sigma},H_\text{imp}\right]c^\dagger_{\sigma}}\nonumber\\
    &\Leftrightarrow \sum_\nu (-\imath\nu) g_{\nu\sigma}e^{\imath\nu\tau} = -\delta(\tau) + \av{T_\tau\left[c_{-\tau,\sigma},H_\text{imp}\right]c^\dagger_{\sigma}}.\label{app:impurity_eom}
\end{align}
The time derivative in the EOM is taken from the left ($-\tau\leq 0$). This is done in order to approach the equal time limit $\tau\rightarrow 0$ without a jump in $g(\tau)$.
To evaluate the RHS, we need to build the commutator $[c_\sigma,H_i]$ with all the components of the impurity Hamiltonian~\eqref{bose_anderson}.
For this we need the commutators
\begin{align*}
  [c^{}_\sigma,c^\dagger_{\sigma'}] &=\delta_{\sigma,\sigma'}-2c^\dagger_{\sigma'}c^{}_\sigma,\\
  [c_\sigma,c_{\sigma'}] &=-2c_{\sigma'}c_\sigma,\\
  [c_\sigma,n_\up n_\dn] &= (n_\up \delta_{\sigma\dn}+n_\dn \delta_{\sigma\up})c_\sigma,\\
  [c_\sigma,\bar{\rho}^\alpha] &=[c_\sigma,\rho^\alpha]= \sum_{\sigma'}s^\alpha_{\sigma\sigma'}c_{\sigma'}.
\end{align*}
Then the commutators on the RHS of the EOM~\eqref{app:impurity_eom} become
\begin{align*}
  \left[c_\sigma,H_\text{at}\right] &= -\tilde{\mu}c_\sigma+\tilde{U} (n_\up \delta_{\sigma\dn}+n_\dn \delta_{\sigma\up})c_\sigma,\\
  \left[c_\sigma,H_\Delta\right] &= \sum_{\vect{k}\sigma'}(v^{}_{\vect{k}}[c_\sigma,c^\dagger_{\sigma'}f^{}_{\vect{k}\sigma'}]+v^*_{\vect{k}}[c_\sigma,f^\dagger_{\vect{k}\sigma'}c^{}_{\sigma'}])\nonumber\\
  &=\sum_{\vect{k}}v^{}_{\vect{k}}f^{}_{\vect{k}\sigma},\\
  \left[c_\sigma,H^\alpha_\Lambda\right]&=\sum_{\vect{q}}w^\alpha_{\vect{q}}\phi^\alpha_{\vect{q}}\sum_{\sigma'}s^\alpha_{\sigma\sigma'}c_{\sigma'}.\notag
\end{align*}
The other components of $H_\text{imp}$ commute with $c_\sigma$. We insert these results into~\eqref{app:impurity_eom} and sum over $\sigma$ on both sides:
\begin{align}
 &\sum_\sigma(-\partial_\tau) g_{-\tau\sigma}=-2\delta(\tau) -\tilde{\mu}\sum_\sigma\av{T_\tau c_{-\tau,\sigma}c^\dagger_{\sigma}}\nonumber\\
 +&\tilde{U}\sum_\sigma\av{T_\tau n_{-\tau,-\sigma}c^{}_{-\tau\sigma} c^\dagger_\sigma}+\sum_{\vect{k}\sigma}v^{}_{\vect{k}}\av{T_\tau f^{}_{-\tau\vect{k}\sigma}c^\dagger_\sigma}\nonumber\\
  &+\sum_{\alpha\vect{q}}w^\alpha_{\vect{q}}\av{T_\tau \phi^\alpha_{-\tau\vect{q}}\sum_{\sigma\sigma'}s^\alpha_{\sigma\sigma'}c_{-\tau\sigma'}c^\dagger_\sigma}.
\end{align}
We identify the impurity Green's function $g_{-\tau\sigma}=-\av{T_\tau c_{-\tau,\sigma}c^\dagger_{\sigma}}$ on the RHS and order the remaining expectation values by time ($-\tau<0$).
\begin{align}
    &\sum_{\sigma}(-\partial_\tau-\tilde{\mu}) g_{-\tau\sigma}\notag\\
  = &-2\delta(\tau) -\tilde{U}\sum_\sigma\av{T_\tau c^\dagger_\sigma n_{-\sigma}c^{}_{-\tau\sigma}}-\sum_{\vect{k}\sigma}v^{}_{\vect{k}}\av{T_\tau c^\dagger_\sigma f^{}_{-\tau\vect{k}\sigma}}\nonumber\\
    &-\sum_{\alpha\vect{q}}w^\alpha_{\vect{q}}\av{T_\tau \phi^\alpha_{-\tau\vect{q}}\sum_{\sigma\sigma'}s^\alpha_{\sigma\sigma'}c^\dagger_\sigma c_{-\tau\sigma'}}.\label{app:docc_ordered}
\end{align}
The aim is to take the EOM at $\tau=0$, we insert $0^+$ in places where the limit is dubious.
Time-ordered products sort creation operators to the left of annihilators at equal time.
Since we approach this point from $-\tau<0$, the order of operators in Eq.~\eqref{app:docc_ordered} remains unchanged in the limit, avoiding jumps in the expectation values.
At equal time we recognize the double occupancy $\av{c^\dagger_\sigma n_{-\sigma}c^{}_{\sigma}}=\av{n_{-\sigma}n_{\sigma}}=d$ and the density operators $\sum_{\sigma\sigma'}s^\alpha_{\sigma\sigma'}c^\dagger_\sigma c_{\sigma'}=\rho^\alpha$.
The averages $\sum_{\vect{k}}v^{}_{\vect{k}}\av{c^\dagger_\sigma f^{}_{\vect{k}\sigma}}=\sum_{\nu}\Delta_{\nu\sigma}g_{\nu\sigma}$
and $\sum_{\vect{q}}w^\alpha_{\vect{q}}\av{\phi^\alpha_{\vect{q}}\bar{\rho}^\alpha}=-\sum_{\omega}\tilde{\Lambda}^\alpha_\omega\chi^\alpha_\omega$ have been obtained from functional derivatives in the previous section.
To use the latter in Eq.~\eqref{app:docc_ordered}, one has to account for the difference in $\rho$ and $\bar{\rho}=\rho-\av{\rho}$.
Back in Eq.~\eqref{app:docc_ordered} we have
\begin{align}
  &-\sum_{\nu\sigma}(\imath\nu+\tilde{\mu}-\Delta_{\nu\sigma}) g_{\nu\sigma}e^{\imath\nu 0^+}\\
 =& -2\delta(0^+) -2\tilde{U}d+\sum_{\alpha\omega}\tilde{\Lambda}^\alpha_\omega \chi^\alpha_\omega-\tilde\Lambda^\alpha_{\omega=0}\av{\rho^\alpha}^2.\nonumber
\end{align}
We recognize $g^0=[\imath\nu+\tilde\mu-\Delta]^{-1}$ on the LHS and use Dyson's equation, $g/g^0=1+\Sigma g$.
$\sum_{\nu\sigma}e^{\imath\nu 0^+}=2\delta(0^+)$ on the LHS cancels, leaving us with a modified Migdal-Galitzkii formula,
\begin{align}
    d=\frac{1}{2\tilde{U}}\sum_{\nu\sigma}g_{\nu\sigma}\Sigma_{\nu\sigma}e^{\imath\nu 0^+}+\frac{1}{2\tilde{U}}\sum_{\alpha\omega}\tilde{\Lambda}^\alpha_\omega (\chi^\alpha_\omega-\beta\av{\rho^\alpha}^2\delta_\omega).\label{app:modified_migdal}
\end{align}
The factor $\beta$ on the RHS accounts for the factor $\beta^{-1}$ implied in the frequency summation.
The above formula does not express the double occupancy $d$ in terms of 1P quantities, due to the contribution of $\chi$ on the RHS.
However, bringing this contribution to the LHS, one still obtains a relation between 2P and 1P quantities,
which justifies the label $d^{\text{1P}}_\text{imp}$ used in the main text.

To accurately calculate the sum $\sum_{\nu\sigma}g_{\nu\sigma}\Sigma_{\nu\sigma}e^{\imath\nu 0^+}$, one needs to separate the constant Hartree part from the self-energy $\Sigma_\nu=\Sigma_\nu'+\Sigma^\text{H}$
and to treat the asymptotes of $g_\nu=1/\imath\nu + ...$ and $\Sigma_\nu'=c_1/\imath\nu + ...$ analytically. Then, the equal time limit can be taken safely.
\begin{align}
  &\sum_{\nu\sigma}g_{\nu\sigma}\Sigma_{\nu\sigma}e^{\imath\nu 0^+}=\sum_{\nu\sigma}g_{\nu\sigma}(\Sigma_{\nu\sigma}'+\Sigma^\text{H})e^{\imath\nu 0^+}\\
  &=\sum_{\nu\sigma}(g_{\nu\sigma}\Sigma_{\nu\sigma}'-c_1/(\imath\nu)^2)-c_1\beta/2+\av{n}\Sigma^\text{H}.
\end{align}

\section{Ward identities and asymptotes in the Bose-Fermi-Kondo model}
\label{app:Wardlocal}
We derive the Ward identities of the Bose-Fermi-Kondo model~\eqref{bose_anderson} and establish a relation to the susceptibility asymptotes derived in Appendix~\ref{app:asymptotics:impurity}.
\subsection{Ward identities}
\label{app:WardBFK}
We can follow the derivation for the Ward identities on the lattice in Appendix~\ref{app:Wardlat} which led to~\eqref{app:ward_intermediate}. Omitting momentum indices, 
and inserting impurity instead of lattice quantities, $G_k\rightarrow g_\nu$, $G^{(2),\alpha}_{kk'q}\rightarrow g^{(2),\alpha}_{\nu\nu'\omega}$, $X^\alpha_{kk'q}\rightarrow\chi^\alpha_{\nu\nu'\omega}$, $\Gamma^\alpha_{kk'q}\rightarrow\gamma^\alpha_{\nu\nu'\omega}$, $H\rightarrow H_{\text{imp}}$, we obtain in frequency space
\begin{align}
&\frac{1}{2}\sum_{\sigma\sigma'}s^\alpha_{\sigma'\sigma}\left\{-\imath\omega\langle c^{}_{\nu\sigma}c^{\dagger}_{\nu+\omega,\sigma'}\rho^\alpha_{\omega}\rangle
+\langle c^{}_{\nu\sigma}c^{\dagger}_{\nu+\omega,\sigma'}[\rho^\alpha_\omega,H_{\text{imp}}]\rangle\right\}\notag\\
&=g_{\nu+\omega}-g_\nu.\label{app:ward_general_local}
\end{align}
The impurity Hamiltonian $H_{\text{imp}}$ is defined in~\eqref{bose_anderson}, where $[\rho^\alpha_\omega,H_{\text{at}}]=[\rho^\alpha_\omega,H^0_{\Delta}]=[\rho^\alpha_\omega,H^0_{\Lambda}]=0$.
We treat the remaining contributions from $H_{\Delta}$ and $H^\alpha_\Lambda$ separately using
$[\rho^\alpha,H_\Delta]=\sum_{\vect{k}\sigma\sigma'}s^\alpha_{\sigma'\sigma}(v_{\vect{k}} c^\dagger_{\sigma'}f^{}_{\vect{k}\sigma}-v^*_{\vect{k}} f^\dagger_{\vect{k}\sigma'}c^{}_{\sigma})$.
We need to calculate the following correlation function in Eq.~\eqref{app:ward_general_local},
\begin{align}
&\frac{1}{2}\sum_{\sigma\sigma'}s^\alpha_{\sigma'\sigma}\langle c^{}_{\nu\sigma}c^{\dagger}_{\nu+\omega,\sigma'}[\rho^\alpha_\omega,H_{\Delta}]\rangle\notag\\
=&\frac{1}{2}\sum_{\sigma_1\sigma_1'\sigma_2\sigma_2'} s^\alpha_{\sigma_1'\sigma_1}s^\alpha_{\sigma_2'\sigma_2}
\left\{\sum_{\vect{k}\nu'}v_{\vect{k}}\langle c^{}_{\nu\sigma_1}c^{\dagger}_{\nu+\omega,\sigma_1'}c^\dagger_{\nu'\sigma_2'}f^{}_{\vect{k}\nu'+\omega,\sigma_2}\rangle\right.\notag\\
-&\left.\sum_{\vect{k}\nu'}v^*_{\vect{k}}\langle c^{}_{\nu\sigma_1}c^{\dagger}_{\nu+\omega,\sigma_1'}f^\dagger_{\vect{k}\nu'\sigma_2'}c^{}_{\nu'+\omega\sigma_2}\rangle\right\}.
   \label{app:ward_bath_raw}
\end{align}
As demonstrated in Appendices~\ref{app:asymptotics:impurity} and~\ref{app:migdal}, the bath operators in the impurity averages can be transferred to Grassmann numbers in the path integral formalism and can then be integrated out.
This leaves convolutions with the hybridization function $\Delta_{\nu\sigma}=\sum_{\vect{k}}|v_{\vect{k}}|^2\mathcal{G}_{\vect{k}\sigma\nu}$, cf.~\eqref{discretedelta}.
We ascertain that the following replacements are valid: "$\sum_{\vect{k}}v_{\vect{k}}f_{\nu\sigma}\rightarrow\Delta_{\nu\sigma}c_{\nu\sigma}$" and
"$\sum_{\vect{k}}v^*_{\vect{k}}f^*_{\nu\sigma}\rightarrow\Delta_{\nu\sigma}c^*_{\nu\sigma}$".
Performing these replacements in Eq.~\eqref{app:ward_bath_raw} and identifying the four-point correlation function,
\begin{align}
g^{(2),\alpha}_{\nu\nu'\omega}=-\frac{1}{2}\sum_{\sigma_1\sigma_1'\sigma_2\sigma_2'}s^\alpha_{\sigma_1'\sigma_1}s^\alpha_{\sigma_2'\sigma_2}
\langle{c^{}_{\nu\sigma_1}c^{\dagger}_{\nu+\omega,\sigma_1'}c^{}_{\nu'+\omega,\sigma_2}c^{\dagger}_{\nu'\sigma_2'}}\rangle,\notag
\end{align} 
and the generalized susceptibility $\chi^\alpha_{\nu\nu'\omega}=g^{(2),\alpha}_{\nu\nu'\omega}+2\beta g_\nu g_{\nu'}\delta_\omega\delta_\alpha$ of the impurity,
one can separate the non-interacting part of~\eqref{app:ward_general_local} as 
\begin{align}
    g_{\nu+\omega}-g_\nu=&\sum_{\nu'}\chi^{\alpha}_{\nu\nu'\omega}\left[\Delta_{\nu'+\omega}-\Delta_{\nu'}-\imath\omega\right]\notag\\
  &+\frac{1}{2}\sum_{\sigma\sigma'}s^\alpha_{\sigma'\sigma}\langle c^{}_{\nu\sigma}c^{\dagger}_{\nu+\omega,\sigma'}[\rho^\alpha_\omega,H_{\Lambda}]\rangle.\label{app:ward_general_impurity}
\end{align}
We are left with the commutator $[\rho^\alpha,H_{\Lambda}]$ on the RHS with the retarded interactions $H_{\Lambda}=\sum_\alpha H^\alpha_{\Lambda}$.
We recall that $H_{\Lambda}=\sum_{\alpha\vect{q}}w^\alpha_{\vect{q}}\bar{\rho}^\alpha\phi^\alpha_{\vect{q}}$ and recognize that $[\rho^0,H_{\Lambda}]=0$,
that is, the retarded interactions do not contribute to the charge current.
$\Lambda^{x,y,z}$ on the other hand do contribute to the spin currents due to the commutation relations for spin operators,
$[\rho^\alpha,\rho^\beta]=2\imath\sum_\gamma\varepsilon_{\alpha\beta\gamma}\rho^\gamma$,
\begin{align}
    [\rho^\alpha,H_{\Lambda}]= 2\imath\sum_{\beta\gamma}\varepsilon_{\alpha\beta\gamma}\rho^\gamma\sum_{\vect{q}}w^\beta_{\vect{q}}\phi^\beta_{\vect{q}}.
\end{align}
Inserting this relation into Eq.~\eqref{app:ward_general_impurity} one can transfer the bosonic operators $\phi^\beta$ into complex variables in the path integral and integrate these out.
As exercised in Appendix~\ref{app:asymptotics:impurity}, this leads to a replacement rule, "$\sum_{\vect{q}}w^\beta_{\vect{q}}\phi^\beta_\omega\rightarrow\tilde{\Lambda}^\beta_\omega\rho^\beta_\omega$",
and hence $\phi^\beta$ give rise to the retarded spin-spin interaction $\tilde{\Lambda}^\beta$. We obtain a six-point correlation function on the RHS of Eq.~\eqref{app:ward_general_impurity},
leaving us with the Ward identities of the BFK,
\begin{align}
    &g_{\nu+\omega}-g_\nu=\sum_{\nu'}\chi^{\alpha}_{\nu\nu'\omega}\left[\Delta_{\nu'+\omega}-\Delta_{\nu'}-\imath\omega\right]\label{app:ward_bfk}\\
    &+\frac{1}{2}\sum_{\sigma\sigma'}s^\alpha_{\sigma'\sigma}\sum_{\omega'\beta\gamma}2\imath\varepsilon_{\alpha\beta\gamma}\tilde{\Lambda}^\beta_{\omega'}
 \langle c^{}_{\nu\sigma}c^{\dagger}_{\nu+\omega,\sigma'}\rho^\beta_{\omega'}\rho^\gamma_{\omega-\omega'}\rangle.\notag
\end{align}
The six-point correlation function in the second line of Eq.~\eqref{app:ward_bfk} contains essentially 3-particle irreducible contributions
(which cannot be broken up into parts by cutting one, two or three fermion lines).
Hence the Ward identities of the BFK~\eqref{app:ward_bfk} cannot be recast into a relation
which features only the one- and two-particle-irreducible vertices $\Sigma$ and $\gamma^{\alpha}$.

Assuming an isotropic retarded interaction $\tilde{\Lambda}^{x,y,z}=\tilde{\Lambda}'$,
we can sum the impurity Ward identities~\eqref{app:ward_bfk} over $\nu$ and $\alpha$,
use the definition of the spin-density operators $\rho^{x,y,z}=2S^{x,y,z}=\sum_{\sigma\sigma'}c^\dagger_{\sigma}s^\alpha_{\sigma\sigma'}c_{\sigma'}$
and the definition of the vector product, $(\mathbf{A}\times\mathbf{B})^\alpha=\sum_{\beta\gamma}\varepsilon_{\alpha\beta\gamma}A^\beta B^\gamma$,
to see that $\tilde{\Lambda}'$ couples to the time-dependent spin-chirality~\cite{Wen89},
\begin{align}
    &\sum_{\nu\nu'\alpha}\chi^{\alpha}_{\nu\nu'\omega}\left[\Delta_{\nu'+\omega}-\Delta_{\nu'}-\imath\omega\right]\notag\\
   =&8\imath\sum_{\omega'}\tilde{\Lambda}_{\omega'}'\langle\mathbf{S}_{-\omega}(\mathbf{S}_{\omega'}\times\mathbf{S}_{\omega-\omega'})\rangle.
    \label{app:spin_chirality}
\end{align}
Writing the RHS in imaginary time, 
$\int_0^\beta\tilde{\Lambda}_{\tau_2-\tau_3}'\langle T_\tau[\mathbf{S}_{\tau_1}(\mathbf{S}_{\tau_2}\times\mathbf{S}_{\tau_3})]\rangle d\tau_2$,
the spin-chirality obviously vanishes when two of its time-indices are equal. 
Hence this contribution arises exclusively in presence of a time-dependent spin-spin interaction.

\begin{figure}[h]
    \vspace{0.1cm}
\begin{tikzpicture}
       \newdimen\R
       \R=1.cm
             \begin{scope}[thick,decoration={
                             markings,
                                 mark=at position 0.5 with {\arrow{<}}}
                                     ]
          \draw (30:\R)
             \foreach \x in {90,150,...,330,30} {  -- (\x:\R) };
                \draw [black,postaction={decorate},>=latex] (30:\R) to [bend left=10] ($(0:\R)+(1.2,0)$) ;
                \draw [black,postaction={decorate},>=latex] ($(0:\R)+(1.2,0)$) to [bend right=30] ($(90:\R)$);
       
                \draw [black,postaction={decorate},>=latex] (270:\R) to [bend right=30] ($(0:\R)+(1.2,0)$) ;
                \draw [black,postaction={decorate},>=latex] ($(0:\R)+(1.2,0)$) to [bend left=10] ($(330:\R)$);
       
                \draw [black,postaction={decorate},>=latex] (150:\R) to ($(150:\R)+(-.7,0)$) ;
                \draw [black,postaction={decorate},>=latex] ($(210:\R)+(-.7,0)$) to (210:\R);
         \end{scope}

             \draw[fill=black] (0:\R)+(1.2,0) circle (1pt);
             \draw ($(0:\R)+(2.,0)$) node{$\imath\varepsilon_{\alpha\beta\gamma}\tilde{\Lambda}^\beta$} ;
  \draw ($(60:\R*0.6)$) node{$\gamma$} ;
  \draw ($(180:\R*0.6)$) node{$\alpha$} ;
  \draw ($(300:\R*0.6)$) node{$\beta$} ;

\end{tikzpicture}
\caption{Symbolic representation of the second line of Eq.~\eqref{app:ward_bfk}. One obtains a representation of the RHS of Eq.~\eqref{app:spin_chirality} by tapering the open Green's function lines (implying summation over $\alpha=x,y,z$).}
\label{fig:3bosonvertex}
\end{figure}
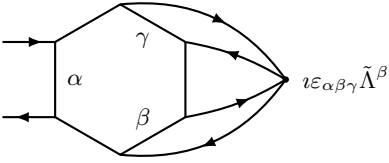
In the Anderson impurity model we have $\Lambda=0$ and the interaction $Un_\uparrow n_\downarrow$ does not contribute to the currents.
In this case the six-point correlation function drops out of the impurity Ward identities~\eqref{app:ward_bfk}.
Then, analogous to the lattice Ward identities~\eqref{app:ward}, one can recast the impurity Ward identities into the form $\Sigma_{\nu+\omega}-\Sigma_\nu = -\sum_{\nu'}\gamma_{\nu\nu'\omega}^{\alpha}[g_{\nu'+\omega}-g_{\nu'}]$, Eq.~\eqref{eq:Wardimp} in the main text.

\subsection{Relation to susceptibility asymptotes}
\label{app:WardAIM}
We prove that the Ward identities of the AIM~\eqref{eq:Wardimp} determine the $(\imath\omega)^{-2}$ coefficient of the impurity susceptibility.
Consequently, finding a different coefficient proves a violation of Eq.~\eqref{eq:Wardimp}, which is used in section~\ref{section:conservation_scdb}.
In the AIM we have $\tilde{\Lambda}=0$. Then, similar considerations as in Appendix~\ref{app:asymptotefromward} [cf. Eq.~\eqref{app:sumrule_largeomega}]
show that the impurity Ward identities~\eqref{app:ward_bfk} imply the following high-frequency asymptote of the susceptibility,
\begin{align}
    \lim\limits_{\omega\rightarrow\infty}(\imath\omega)^2 \chi^\alpha_{\omega}&=\lim\limits_{\omega\rightarrow\infty} 2\sum_{\nu}(g_{\nu}-g_{\nu+\omega})\left[\Delta_{\nu+\omega}-\Delta_{\nu}\right]\nonumber\\
    &=-4\sum_{\nu}g_{\nu}\Delta_{\nu}\label{app:sumrule_largeomega_aim}.
\end{align}
In the last step it was used that $g_{\nu+\omega}, \Delta_{\nu+\omega}$ vanish at $\omega\rightarrow\infty$.
Eq.~\eqref{app:sumrule_largeomega_aim} is in agreement with the asymptotes of $\chi$ in the BFK for $\tilde{\Lambda}^{x,y,z}=0$ [see Eq.~\eqref{app:xyzasymptote}].
For $\tilde{\Lambda}\neq 0$ we find by comparison of Eq.~\eqref{app:sumrule_largeomega_aim}
with the asymptote of the exact solution~\eqref{app:xyzasymptote} that
the 6-point correlation function in the impurity Ward identities~\eqref{app:ward_bfk} contributes to the asymptote of the spin susceptibility $\chi^{\alpha=x,y,z}$
with $4\sum_{\omega',\beta\neq\alpha}\tilde{\Lambda}^\beta_{\omega'} \chi^\beta_{\omega'}$.

\bibliography{main}

\end{document}